\documentclass[12pt, a4paper]{article}

\pdfoutput=1

\usepackage{amsmath}
\allowdisplaybreaks
\usepackage{amsfonts}
\usepackage{amssymb}
\usepackage{bbm}
\usepackage{verbatim}
\usepackage{dsfont}
\usepackage{booktabs}
\usepackage{slashed}
\usepackage{bm}

\usepackage{epsfig}
\usepackage{color}
\usepackage[table]{xcolor}
\usepackage{graphicx}
\usepackage[]{caption}
\usepackage[listofformat=empty,subrefformat=empty]{subfig}

\usepackage{float}
\usepackage{overpic}

\usepackage{enumerate}
\usepackage{hhline}
\usepackage{multirow}

\usepackage{cite}
\usepackage{xspace}
\usepackage{setspace}

\usepackage{bbold}

\usepackage[pdftitle={Effective field theory for magnetic compactifications},
  pdfauthor={},
  pdfsubject={},
  bookmarksopen, bookmarksnumbered, bookmarksopenlevel=2, colorlinks=false, linkcolor=blue, citecolor=blue, urlcolor=blue]{hyperref}

\DeclareMathOperator{\tr}{\text{tr}}

\renewcommand{\tfrac}{\genfrac{}{}{}1}

\newcommand{\bz}{\overline{z}}

\newcommand{\Tr}{\text{Tr}}
\renewcommand{\tr}{\text{tr}}
\newcommand{\dg}{\dagger}

\newcommand{\ophi}{\bar{\phi}}

\newcommand{\E}{\varepsilon}

\newcommand{\pd}{\partial}
\newcommand{\opd}{\overline{\partial}}

\newcommand{\kk}{\eta}
\newcommand{\bk}{\overline{\eta}}
\newcommand{\vf}{\varphi}
\newcommand{\ovf}{\overline{\varphi}}

\begin{document}

\thispagestyle{empty}

\begin{flushright}
CPHT-RR066.092020\\
DESY 20-163\\
\end{flushright}
\vskip .8 cm
\begin{center}
  {\Large {\bf Tachyon condensation  in magnetic compactifications}}\\[12pt]

\bigskip
\bigskip 
{
{\bf{Wilfried Buchmuller$^\dagger$}\footnote{E-mail: wilfried.buchmueller@desy.de}},
{\bf{Emilian Dudas$^\ast$}\footnote{E-mail: emilian.dudas@cpht.polytechnique.fr}},
{\bf{Yoshiyuki Tatsuta$^{\ddag, \, \dagger}$}\footnote{E-mail: yoshiyuki.tatsuta@sns.it}}
\bigskip}\\[0pt]
\vspace{0.23cm}
{\it $^\dagger$ Deutsches Elektronen-Synchrotron DESY, 22607 Hamburg, Germany \\ \vspace{0.2cm}
$^\ast$ CPHT, CNRS, Institute Polytechnique de Paris,  France \\ \vspace{0.2cm}
$\ddag$ Scuola Normale Superiore and INFN, Piazza dei Cavalieri 7, Pisa 56126, Italy}\\[20pt] 
\bigskip
\end{center}

\date{\today{}}

\begin{abstract}
\noindent
Intersecting D-brane models and their T-dual magnetic
compactifications provide an attractive framework for particle
physics, allowing for chiral fermions and supersymmetry breaking.
Generically, magnetic compactifications have tachyons that are
usually removed by Wilson lines. However, quantum corrections prevent 
local minima for Wilson lines. 
We therefore study tachyon condensation in the simplest case,
the magnetic compactification of type I string theory on a torus to
eight dimensions. We find that tachyon condensation restores 
supersymmetry, which is broken by the magnetic flux, and we
compute the mass spectrum of vector- and hypermultiplets.
The gauge group $\text{SO}(32)$ is broken to $\text{USp}(16)$.
We give arguments that the vacuum reached by tachyon
  condensation corresponds to
  the unique 8d superstring theory already known in the literature,
  with discrete
 $B_{ab}$ background or, in the T-dual version, the type IIB
 orientifold with three $\text{O}7_-$-planes, one $\text{O}7_+$-plane and eight
 D7-branes coincident with the  $\text{O}7_+$-plane.
The ground state after tachyon condensation is supersymmetric and has
no chiral fermions.
\end{abstract}

\newpage 
\setcounter{page}{2}
\setcounter{footnote}{0}

{\renewcommand{\baselinestretch}{1.5}

\section{Introduction}
\label{sec:Introduction}

Intersecting D-brane models and their T-dual magnetic
compactifications remain an attractive framework for
string theory compactifications to four dimensions with chiral fermion spectra
\cite{Ibanez:2012zz}. The rich structure of D-branes and fluxes
provides an intuitive way to explore the structure of string vacua
with unbroken as well as broken supersymmetry
\cite{Angelantonj:2002ct,Blumenhagen:2006ci}.
In the absence of any hint for supersymmetry at the Large Hadron Collider,
extensions of the Standard Model where supersymmetry is broken at a
high scale are of particular interest, and magnetic string compactifications
provide a theoretical framework for phenomenological models like
`split supersymmetry' \cite{ArkaniHamed:2004fb,Giudice:2004tc,
  Antoniadis:2004dt} or `split symmetries' \cite{Buchmuller:2015jna}. 

An intriguing aspect of magnetic compactifications is the connection
between fermion chirality and supersymmetry breaking
\cite{Bachas:1995ik}, which occurs in
compactifications of type I strings on tori and orbifolds
\cite{Abouelsaood:1986gd,Blumenhagen:2000wh,Angelantonj:2000hi},
and correspondingly in the related intersecting D-brane models
\cite{Berkooz:1996km,Blumenhagen:2000wh,Aldazabal:2000dg}.
In this framework compactifications have been constructed
that come rather close to the Standard Model of
particle physics \cite{Aldazabal:2000cn,Ibanez:2001nd,Blumenhagen:2001te,Cvetic:2001nr}. 
A generic feature of models with broken supersymmetry is the
appearance of tachyons. In all known orientifold constructions with
broken supersymmetry they are removed by
means of Wilson lines that correspond  flat directions in the tree-level potential.
However, in a representative example, we recently showed that quantum
corrections render these vacua unstable and that the theory is always
driven to the tachyonic regime \cite{Buchmuller:2019zhz}, which represents a serious challenge
to the known models.

In principle, tachyonic instabilities are not necessarily a problem, and tachyon
condensation may even lead to vacua that are phenomenologically
welcome. However, the process of tachyon condensation is complicated
\cite{Hashimoto:2003xz,Epple:2003xt, Sen:2004nf}. In the following we
therefore study the simplest case in some depth, namely type I string compactification
on a magnetized torus to eight dimensions. The magnetic flux breaks
the symmetry group $\text{SO}(32)$ to $\text{U}(16)$ and produces two tachyons. 
We find that their condensation breaks $\text{U}(16)$ further to $\text{USp}(16)$
and restores supersymmetry, which is broken by the
magnetic flux. We also 
compute the mass spectrum of massive vector- and hypermultiplets.
Remarkably, this turns out to be consistent with
the unique 8d superstring theory that is already known in the literature.
It has gauge group $\text{USp}(16)$ and contains
a discrete $B_{ab}$ flux background. In the T-dual version, this
corresponds to the type IIB orientifold with three $\text{O}7_-$-planes and one $\text{O}7_+$-plane, with eight D7-branes required
by tadpole cancellation coincident with the $\text{O}7_+$-plane.

The paper is organized as follows. In Section~\ref{sec:magnetic} we
briefly discuss the magnetic compactification to eight dimensions,
its partition function, the tachyonic instability and the effective
field theory in eight dimensions. The symmetry breaking to
$\text{USp}(16)$ is explained in Section~\ref{sec:symmetry}. 
Moreover, a Wilson line is considered that breaks $\text{SO}(32)$ to
$\text{U}(16)$ rather than $\text{USp}(16)$, leaving
supersymmetry unbroken. Tachyon condensation is discussed
in Section~\ref{sec:singular} and shown to lead to singular localized
flux. Supersymmetry is unbroken and there are no chiral fermions.
As shown in Section~\ref{sec:massive},
the mass spectrum corresponds to a free theory with twisted boundary conditions.
In Section~\ref{sec:d9d7} a brane description of the vacuum after tachyon
condensation is described, which is shown to correspond to the unique
8d $\text{USp}(16)$ in Section~\ref{sec:unique}. In the appendix,
for the convenience of the reader, we recall flux quantization on the
torus in the symmetric gauge.


\section{Magnetic compactification to eight dimensions}
\label{sec:magnetic}

\subsection{Intersecting D8-branes}

Let us consider the simplest system in which tachyon condensation
occurs: intersecting D8-branes in type IIA string theory compactified
on a toroidal orientifold with two compact dimensions, a setup
very similar to the magnetic compactifications to six dimensions 
considered in \cite{Blumenhagen:2000wh}. Here
the 10d theory is compactified on a torus to 8d
Minkowski space with real coordinates $x_0, \ldots ,
x_3$ and complex coordinates $z_i = (x_{2+2i} + i x_{3+2i})/2$,
$i=1,2$. The complex torus coordinate is $z\equiv z_3$, with the
identifications $z \sim z + \pi R$, 
$z \sim z +i \pi R'$. An orientifold is obtained by dividing
out the discrete symmetry $\Omega\mathcal{R}(-1)^{F_L}$,
where $\Omega$ is worldsheet parity, $F_L$ is left-moving
fermion number, and $\mathcal{R}$ is the symmetry of $T^2$ under the reflection
$z \rightarrow \bar{z}$. The orientifold has two O8-planes along 8d
Minkowski space, and the $x_{8}$-direction is invariant under $\mathcal{R}$. The
orientifold planes are localized at the fixed points $x_9 = 0,
\pi R'$.  Each orientifold plane has RR charge $Q_{\text{O}8} = -16$ in
units of a D8-brane charge.
A stack of $N$ D8-branes requires a stack of $N$ mirror D8-branes to
satisfy the reflection symmetry $\mathcal{R}$ of the
compact space. The branes are wrapped around the  1-cycles $[a]$
and $[b]$ of the 2-torus with wrapping numbers $n$ and $m$,
corresponding to the homology class $[\Pi] = n [a] + m[b]$.
The homology class $[\Pi]'$ of the mirror brane is obtained from
$[\Pi]$ by replacing $m$ by $-m$.
The stack of $N$ branes carries the gauge symmetry $\text{U}(N)$, and the RR
tadpole cancellation condition reads
\begin{equation}\label{RRcancellation}  
 N ([\Pi] + [\Pi]')  -  16 [\Pi_{\text{O}8}] = 0\ .
\end{equation}
For simplicity we restrict ourselves to the case $n=1$.
With $[\Pi_{\text{O}8}] = 2 [a]$, this implies $N=16$.

The stack of branes leads to a tower of 8d $\mathcal{N}=1$
supermultiplets in the adjoint representation ${\bf 256}$ of the gauge group
$\text{U}(16)$. Branes and mirror branes intersect at an angle determined by
the wrapping numbers. The wrapping numbers we consider here are $(n,m)=(1,k)$ for the D8 branes and $(1,-k)$ for their mirrors. At their intersection a massless chiral 8d $\mathcal{N}=1$
fermion  in the antisymmetric representation ${\bf 120}$ of $\text{U}(16)$ is
localized, together with massive fermions, vectors, scalars and tachyons, all in the
antisymmetric representation. Their masses depend on the intersection
angle $\theta$ that breaks supersymmetry, and their multiplicity is
given by the intersection number $I$ that is related to the 
wrapping number $k$ by \cite{Blumenhagen:2000wh}
\begin{equation}
  I = 2 k\ .
\end{equation}  

The scalar masses of the antisymmetric tensors depend on the angle
with the orientifold plane. We restrict ourselves to small angles with respect
to the orientifold plane,
\begin{equation}\label{thetarho}
\tan{\theta} = k \rho \simeq \theta \ ,\quad
\rho = \frac{R_2}{R_1}\ ,
\end{equation}
where $R_1$ and $R_2$ are the two radii of the torus.
In the T-dual picture small angles correspond to large areas of the
dual torus so that we are able to use a field theory approximation to string
partition functions.

At the intersection of the two stacks of branes and mirror branes one then has
three light scalars with masses \cite{Aldazabal:2000dg}
\begin{equation}\label{scalarmasses}
4\pi\alpha' M^2_1 = -2|\theta| \ ,\quad
4\pi\alpha' M^2_2 = 2|\theta| \ , \quad
4\pi\alpha' M^2_3| = 2|\theta|  \ ,
\end{equation}
where $-\pi/2 \leq \theta \leq \pi/2$. Clearly, the model has a
tachyon with negative mass squared $M_1^2 = - k\rho/(2\pi\alpha')$. 

The intersecting D-brane model described above is
T-dual to a type I compactification on a magnetized dual rectangular torus
$T^2$ with the identifications 
\begin{align}
z \sim z + \pi R_1\ , \quad z \sim z + i \pi\alpha'/R_2 \ .
\end{align}
The area of the dual torus is $4\pi^2\alpha'/\rho$, and the angle
$\theta$ between the stack of branes and the orientifold plane 
is related to a magnetic flux density $f$ on the torus $T^2$ \cite{Berkooz:1996km},
\begin{align}
\tan \theta = 2\pi\alpha' g f\ ,
\end{align}
where $g$ is a gauge coupling.
Using Eq.~\eqref{thetarho},
this implies the quantization condition for the flux density $f$,
\begin{equation}\label{Dquantization}
\frac{g}{2\pi}\int_{T^2} f =  k \ .  
\end{equation} 

The mass spectrum of the intersection D8-branes is encoded in the
magnetic deformation of the open string part of the type I partition
function\footnote{For simplicity, we drop the overall factor $1/(2(4\pi^2\alpha')^4)$.}. This is the sum of  annulus amplitude $\mathcal{A}$ and
Moebius amplitude $\mathcal{M}$, which can be compactly expressed in
terms of standard modular functions and $\text{SO}(8)$ characters \cite{Angelantonj:2002ct},
\begin{align}
  \mathcal{A}  &=   \int_0^{\infty} \frac{d \tau_2}{\tau_2^5} \left\{
  N\bar{N}\frac{(V_8-S_8)(0|\tau)}{\eta^8}   \ + \right.   \nonumber\\
  & \left. \hspace{1cm} 2k \left[ \frac{N^2}{2}
  (V_8-S_8)(2 \epsilon\tau|\tau)
  + \frac{\bar{N}^2}{2} (V_8-S_8)(-2 \epsilon\tau |\tau)  \right]
\frac{i}{\theta_1 (2 \epsilon\tau |\tau) \eta^5}  \right\}  \ ,   \label{ann}\\
  \mathcal{\cal M} &= -k  \int_0^{\infty} \frac{d \tau_2}{\tau_2^5}
  \left[ N ({{\hat V}_8-{\hat S}_8})  (2 \epsilon \tau | \tau )  +
   {\bar N} ({{\hat V}_8-{\hat S}_8})  (-2 \epsilon \tau | \tau )  \right]  
\frac{i}{{ {\hat \theta}_1 (2 \epsilon \tau | \tau ) \hat \eta}^5}  \label{moeb}\ ,   
\end{align} 
where for the annulus amplitude $\tau = i \tau_2/2$, whereas for the Moebius
amplitude $\tau = i \tau_2/2+1/2$, and $N = \bar{N} = 16$. The partition function contains the sum
over string oscillator modes and Landau-level modes, whose masses are
inversely proportional to the area of the torus. Their contribution to the
one-loop effective potential $V = - \mathcal{A}$ is obtained by
considering the $\tau_2 \rightarrow \infty$ limit of the modular functions.  
Changing variables from $\tau_2$ to the Schwinger proper time
$t = \pi\alpha'\tau_2$, the result for the second term in
Eq.~\eqref{ann} can be written as \cite{Buchmuller:2019zhz},
\begin{equation}
  V = \int_0^\infty dt V(t) \propto  \int_0^\infty \frac{dt}{t^5}
  \frac{\sinh^4\left(\frac{gft}{2}\right)}{\sinh\left(gft\right)} \ .
 \end{equation} 
The integral has an infrared divergence at $t\rightarrow \infty$, where
the integrand behaves as
\begin{equation}
  V(t) \propto \frac{1}{t^5} e^{gf t}\ .
\end{equation}
This is the effect of the tachyon with negative mass square $M^2_T = -gf$.


\subsection{Effective field theory in eight dimensions}

Neglecting oscillator modes, the open string sector of the type I
superstring becomes the 10d Super-Yang-Mills action with gauge
group $\text{SO}(32)$.
In order to obtain the 8d effective field theory it is convenient to
start from the 10d Super-Yang-Mills action 
expressed in terms of 4d $\mathcal{N}=1$ vector
superfield $V$ and chiral superfields $\phi$ \cite{Marcus:1983wb,
ArkaniHamed:2001tb},
\begin{equation}\label{10daction}
\begin{split}
  S_{10} = \int d^{10} x \bigg\{ \frac{1}{\tilde{k}} \int d^2 \theta\ 
  \Tr &\Big[ \frac{1}{4} \hat{W} \hat{W} + 
\frac{1}{2}\E_{ijk} \hat{\phi}^i\Big(\pd_j\hat{\phi}^k
    +\frac{g}{3\sqrt{2}}[\hat{\phi}^j,\hat{\phi}^k]  \Big) \Big]
+ \text{h.c.}  \\
 + \frac{1}{\tilde{k}} \int d^4 \theta\ \Tr &\Big[\hat{\phi}^{i\dg} \hat{\phi}^i 
+ \sqrt{2}(\pd_i \hat{\phi}^{i\dg} + \opd_i \hat{\phi}^i) \hat{V}  
 -  g [\hat{\phi}^{i\dg}, \phi^i] \hat{V} \\ 
  &+ \Big(\pd_i \hat{V} + \frac{g}{\sqrt{2}} [\hat{V},\hat{\phi}^i]\Big)^\dg 
\Big(\pd_i \hat{V} +\frac{g}{\sqrt{2}} [\hat{\phi}^i,\hat{V}] \Big) \Big] \bigg\}\ .
\end{split}
\end{equation}
Here $\hat{\phi}^i$ are chiral superfields, $i,j,k=1,2,3$, $\hat{V}$ is a vector
superfield and $\hat{W}$ is the corresponding field strength\footnote{We
  use the conventions of \cite{Wess:1992cp}, and we have dropped the
  WZW term that vanishes in WZ gauge, $V^3=0$. Our trace convention is $\Tr
  \left[ \hat{T}_a \hat{T}_b \right] = \tilde{k}\delta_{ab}$, $\tilde{k}=2$.}. The hat denotes
$\text{SO}(32)$ matrices. 10d Minkowski space is described by real coordinates $x_0, \ldots ,
x_3$, complex coordinates $z_i = (x_{2+2i} + i x_{3+2i})/2$,
$i=1,2$ and $z_3\equiv z$; complex derivatives are defined as $\pd_i = \pd/\pd
z_i$, $\opd_i = \pd/\pd \bar{z}_i$. For simplicity, we consider compactification to 8d
Minkowski space on a square torus $T^2$  with 
the identifications $z \sim z + \pi R$, $z \sim z +i \pi R$.

It is convenient to write the $\text{SO}(32)$ matrix valued fields in block
form\footnote{For a recent discussion, see \cite{Buchmuller:2019zhz}.}, such that the
blocks transform with respect to the $\text{U}(16)$ subgroup as adjoint or
antisymmetric representation, respectively, 
\begin{equation}
\begin{split}  
  \hat{V} &=
  \left(\begin{array}{c c}
          V & \sqrt{2}V^+ \\ \sqrt{2}V^- & -V^T
        \end{array}\right)\ , \quad V = V^\dagger\ , \ V^- =
      V^{+\dagger} \ ,\\
  \hat{\phi}^i &=
  \left(\begin{array}{c c}
          \phi^i & \sqrt{2}\phi^{i+} \\ \sqrt{2}\phi^{i-} & -\phi^{iT}
        \end{array}\right) \ .
\end{split}
\end{equation}  
In terms of these fields\footnote{For fields we use the notation
  $V^\dg = \overline{V}$, $V^{+\dg} = \overline{V}^+ = V_-$, $\phi^\dg
  = \ophi$ etc. $\tr$ denotes the trace of $U(16)$ matrices with the
  convention $\tr[T^a T^b] = \delta_{ab}$, see appendix A of \cite{Buchmuller:2019zhz}} the 10d Lagrangian reads
\begin{equation}\label{L10d}
\begin{split}
  \mathcal{L}_{10} =
  \int d^2 \theta \, 
  \tr &\Big[ \frac{1}{4} W_V W_V + \frac{1}{2} W^+ W^- 
  +\E_{ijk} \Big(\frac{1}{2}\phi^i\pd_j\phi^k \\
&+\phi^{i+}\pd_j\phi^{k-} +\frac{g}{3\sqrt{2}}\phi^i\phi^j\phi^k
-\sqrt{2}g \phi^{i+}\phi^{k-}\phi^j \Big) \Big]
+ \text{h.c.}  \\
 + \int d^4 \theta \, \tr &\Big[\ophi^{i} \phi^i  + \ophi^{i+}\phi^{i+}
 + \ophi^{i-} \phi^{i-} \\
 &+ \big(\sqrt{2}(\pd_i \ophi^{i} + \opd_i\phi^i)
- g[\ophi^{i},\phi^i]
- 2g( \ophi^{i-}\phi^{i-}  -  \phi^{i+}\ophi^{i+})\big)V \\
&+\big(\sqrt{2} (\pd_i \ophi^{i-} + \opd_i \phi^{i+}) - 2g (\ophi^{i}\phi^{i+}
 - \phi^i\ophi^{i-})\big)V^- \\
&+ \big(\sqrt{2}(\pd_i \ophi^{i+} + \opd_i \phi^{i-}) - 2g(\ophi^{i+}\phi^i
- \phi^{i-}\ophi^{i})\big)V^+\\
&+\big(\opd_iV+\frac{g}{\sqrt{2}}[V,\ophi^{i}] + \sqrt{2}g(
V^+\ophi^{i+}  - \ophi^{i-}V^- )\big) \\
&\quad \times\big(\pd_iV +\frac{g}{\sqrt{2}}[\phi^i,V] + \sqrt{2}g(
  \phi^{i+}V^- - V^+\phi^{i-} )\big)\\
&+\big(\opd_iV^+- \frac{g}{\sqrt{2}}\big(\ophi^{i}V^+ + V^+\ophi^{iT}
   - V\ophi^{i-} -\ophi^{i-}V^T)\big)\\
&\quad\times
\big(\pd_iV^--\frac{g}{\sqrt{2}}(V^-\phi^i
                                                   +\phi^{iT}V^--\phi^{i-}V-V^T\phi^{i-})\big)\\
&+\big(\opd_iV^- + \frac{g}{\sqrt{2}}(V^-\ophi^{i} + \ophi^{iT}V^-
- \ophi^{i+}V -V^T\ophi^{i+})\big)\\
&\quad\times  \big(\pd_iV^+ +\frac{g}{\sqrt{2}}(\phi^iV^+ +
V^+\phi^{iT} - V\phi^{i+} - \phi^{i+}V^T)\big)\Big] \ .
\end{split}
\end{equation}

Consider now a background field with constant magnetic flux
($\phi^3\equiv \phi$), which breaks $\text{SO}(32)$ to $\text{U}(16)$,
\begin{equation}\label{vev}
  \langle \phi \rangle = \frac{1}{\sqrt{2}} f\bz \, \mathbb{1} \ ,
\end{equation}
and which is a solution of the equation of motion,
\begin{equation}
\pd  (\pd \langle \ophi \rangle + \opd \langle\phi\rangle) = 0 \ .
\end{equation}
The background field leads to Landau-level type mass terms for charged
fields. They are most easily obtained by combining the magnetic flux
with derivatives to annihilation and creation operators \cite{Buchmuller:2018eog},
\begin{equation}
  \begin{split}
    \pd - gf\bz &= -i\sqrt{2gf}a_-^\dg\ , \quad
\opd + gf z = -i\sqrt{2gf}a_-\ , \\
\opd - gfz &= -i\sqrt{2gf}a_+^\dg\ , \quad
\pd + gf\bz = -i\sqrt{2gf}a_+\ ,
\end{split}    
 \end{equation} 
which satisfy the commutation relations
\begin{equation}
[a_\pm,a^\dagger_\pm]=1\ , \quad [a_\pm,a_\mp]=0\ , \quad
[a_\pm,a^\dagger_\mp]=0\ . 
\end{equation}
The relevant linear and bilinear terms of the Lagrangian \eqref{L10d}
are ($i,j=1,2$, $\phi^{3\pm}=\phi^\pm$)
\begin{equation}\label{10danncrea}
  \begin{split}
\mathcal{L}_{10} \supset \tr\Big[  \int d^2 \theta 
&\ i\sqrt{2gf}\E_{ij}\phi^{i+} a_-^\dg\phi^{j-} + \text{h.c.}  \\
+ \int d^4\theta &\Big(2\big(f - g(\ophi^{i-}\phi^{i-} + \ophi^-\phi^-
- \phi^{i+}\ophi^{i+} - \phi^+\ophi^+)\big)V \\
&-2i\sqrt{gf}\big( (a_-\phi^- + a_-^\dg\ophi^+)V^+
+ ( a_+\ophi^- + a_+^\dg\phi^+)V^-\big)\\
&-2gf \big(a_+^\dg V^+ a_-^\dg V^- + a_- V^- a_+ V^+\big)\Big] \ .
 \end{split}
\end{equation}
Expanding the charged fields in the standard harmonic oscillator mode
functions,
\begin{equation}\label{LL}
\begin{split}
\phi^+ &= \sum_{n,j} \phi^+_{n,j}\xi_{n,j}\ ,\quad \phi^- = \sum_{n,j}
\phi^-_{n,j}\overline{\xi}_{n,j}\ ,  \quad \ophi^+ = \sum_{n,j}
\ophi^+_{n,j}\overline{\xi}_{n,j} \ , \\
\ophi^- &= \sum_{n,j} \ophi^-_{n,j}\xi_{n,j}\ , \quad V^+ = \sum_{n,j}
V^+_{n,j}\xi_{n,j}\ , \quad V^- = \sum_{n,j} V^-_{n,j}\overline{\xi}_{n,j} \ ,
\end{split}
\end{equation}
which satisfy
\begin{equation}\label{axi}
\begin{split}
a_+ \xi_{n,j} &= i \sqrt{n}\ \xi_{n-1,j}\,, \quad a_+^\dagger \xi_{n,j}
= -i\sqrt{n+1}\ \xi_{n+1,j}\,,\\
a_- \overline{\xi}_{n,j} &= i \sqrt{n}\ \overline{\xi}_{n-1,j}\,, \quad a_-^\dagger \overline{\xi}_{n,j}
= -i\sqrt{n+1}\ \overline{\xi}_{n+1,j}\ ,
\end{split}
\end{equation}
keeping only the zero mode of $V$, and integrating over the torus
coordinates, one obtains\footnote{In the following we drop the index
$j$ that  labels the ground state degeneracy.  The multiplicity of
states is given by $qg/(2\pi)\int_{T^2} f = qk$ where $q$ is the
charge of the field. Since the charge of
the antisymmetric tensor is $q=2$, the multiplicity is $2k$
for $k$ flux quanta.}} 
\begin{equation}\label{L8d1}
  \begin{split}
\mathcal{L}_8 \supset \tr\Big[\int d^2 \theta 
& \sum_n\sqrt{2gf(n+1)} \E_{ij}\phi^{i+}_{n+1} \phi^{j-}_{n} + \text{h.c.}  \\
+ \int d^4\theta & \big( \big(2f 
- 2g \sum_n(\ophi^{i-}_{n}\phi^{i-}_{n} - \phi_n^{i+} \phi_n^{i+})\big)V \\
&+2\sqrt{gf}\big( (\sqrt{n+1}\phi^-_{n+1} - \sqrt{n}\ophi^+_{n-1})V^+_n
+ \text{h.c.}\big)\\
&-2gf(2n+1) V^+_n V^-_n \big) \Big] \ .
 \end{split}
\end{equation}
Magnetic flux breaks supersymmetry. To find the mass spectrum one
therefore has to expand superfields in components\footnote{We use the
  same symbol for a chiral superfield and its scalar component.},
\begin{align}
\phi = (\phi,\psi,F) \ , \quad V = (A_\mu,\lambda,D) \ .
\end{align}
The mixing term between chiral and vector superfields contains
D-terms and derivative couplings between Goldstone bosons and
vector fields,
\begin{align}
\sqrt{2}\int d^4 \theta
&\big(\sqrt{n} \ophi^+_{n-1} 
- \sqrt{n+1}\phi^{-}_{n+1}\big)V^+_{n}
\nonumber\\
=&\frac{1}{\sqrt{2}}\big(\sqrt{n} \ophi^+_{n-1} 
- \sqrt{n+1} \ophi^{-}_{n+1}\big)D^+ 
-i\sqrt{2n+1}\ \pd_\mu \Pi^-_{n} A^{+\mu}_{n}\ .
\end{align}
Here the Goldstone fields $\Pi^-$ and the orthogonal complex scalars $\Sigma^-$, formed
from the complex scalars $\ophi^+$ and $\phi^-$, are given by
\begin{equation}
  \begin{split}
\Pi^-_{n} &= \frac{1}{\sqrt{2(2n+1)}}
\big(\sqrt{n} \ophi^+_{n-1} + \sqrt{n+1} \phi^{-}_{n+1}\big) \ ,\\
\Sigma^-_{n} &= \frac{1}{\sqrt{2(2n+3)}}
\big(\sqrt{n+2} \ophi^+_{n} - \sqrt{n+1} \phi^{-}_{n+2}\big) \ .
\end{split}
\end{equation}
The vector bosons of the tower of Landau levels acquire their mass by
the Stueckelberg mechanism, and a shift of the vector bosons, 
\begin{align}
A_{n}^{-\mu} \rightarrow A_{n}^{-\mu} 
+ \frac{i}{\sqrt{gf(2n+1)}}\ \pd_\mu\Pi^{-}_{n}\ ,
\end{align}
cancels the mixings with the Goldstone bosons as well as the kinetic
terms of the Goldstone bosons. Finally, eliminating all F- and D-terms via
their equations of motion, one obtains the bosonic mass terms ($i=1,2$)
\begin{equation} \label{L8db}
  \begin{split}
\mathcal{L}^b_8 = - gf \tr\big[&\sum_{n} 
(2n+1) (A^{+}_{n \mu}A^{-\mu}_{n} + |\phi^{i+}_{n}|^2 +
|\phi^{i-}_{n}|^2)\\
&- |\phi^-_{0}|^2 + \sum_{n} (2n+3) |\Sigma^-_{n}|^2\big] \ .
\end{split}
\end{equation}
The structure of this result is easily understood. The fields
$\phi^{i\pm}_n$ do not mix with the vector bosons, and combining F- and
D-term contributions, one finds that they get the same mass as the vector bosons.
The fields $\phi^\pm_n$ contain the Goldstone bosons for the tower of
vector bosons $A^{\pm\mu}_n$ and an orthogonal tower of fields $\Sigma^-_n$.
Since the vectors mix with different levels of the complex scalars,
the fields $\Sigma^-_n$ are not mass degenerate with $\phi_n^{i\pm}$.
Moreover, the scalar $\phi^-_0$ is a tachyon.\footnote{In
  \cite{Buchmuller:2019zhz} the
  case with magnetic flux in two tori was considered. The mixing
  pattern is then more complicated, and by tuning the fluxes one can
  obtain two massless scalars instead of a tachyon \cite{Aldazabal:2000dg}. The masses
in Eq.~\eqref{L8db} refer to mass eigenstates, contrary to the mass
levels in Bachas' string mass formula \cite{Bachas:1995ik}. There the longitudinal
components of the vectors and the tachyon would be counted as a tower
of scalars with masses $gf(2n-1)$, $n \geq 0$.}

The fermion masses are easily read off from \eqref{L8d1}.
Denoting the Weyl fermions contained in the superfields $\phi^{i\pm}$,
$\phi^\pm$ and $V^+ = \overline{V}^-$ by $\psi^{i\pm}$, $\psi^\pm$
and $\lambda^\pm$, respectively, one obtains
\begin{equation}\label{L8df}
\mathcal{L}_8^f \supset 
\sqrt{gf} \sum_n \sqrt{(n+1)} \tr\big[\E_{ij}\psi^{i+}_{n+1}
\psi^{j-}_{n} + i\psi^-_{n+1}\lambda^+_n + i\psi^+_n\lambda^-_{n+1}\big]
+ \text{h.c.}  
\end{equation}
Again, the mass terms couple different levels of the Landau towers,
and the four 4d Weyl fermions 
$\psi^{1+}_0$, $\psi^{2+}_0$, $\bar{\psi}^-_0$ and $\bar{\lambda}^-_0$
form one 8d Weyl spinor with charge $+1$.
All fields are antisymmetric tensors of $\text{U}(16)$ and have a multiplicity
$2k$ for $k$ flux quanta.

The negative mass squared of $\phi^-_0$ in Eq.~\eqref{L8db} suggests
that this mode will grow exponentially with time after a small
perturbation around the unstable extremum $\phi^-_0 = 0$. At present
it is an open question what the final state after tachyon condensation
is like. Of particular interest are supersymmetric vacua which can be
analyzed by means of auxiliary fields, allowing for background flux as
well as Wilson lines. From Eq.~\eqref{L10d} one obtains
\begin{align}\label{auxiliary}
-\overline{F}^i &=
 \E_{ijk} \big(\pd_j\phi^k 
+\frac{g}{\sqrt{2}}\phi^{j}\phi^{k} -\sqrt{2}g
                  \phi^{j+}\phi^{k-}\big)\ ,\\
-\overline{F}^{i+} &=
 \E_{ijk} \big(\pd_j\phi^{k-} -\sqrt{2}g \phi^{j-}\phi^k \big)\ ,\\
-\overline{F}^{i-} &=
 \E_{ijk} \big(\pd_j\phi^{k+}  -\sqrt{2}g \phi^{j+}\phi^{k}\big)\ , \\
-2D &=  \sqrt{2}(\pd_i \ophi^{i} + \opd_i\phi^i)
- g[\ophi^{i},\phi^i]
     - 2g( \ophi^{i-}\phi^{i-}  -  \phi^{i+}\ophi^{i+}) \label{D0}\ ,\\
-D^+& =  \sqrt{2} (\pd_i \ophi^{i-} + \opd_i \phi^{i+}) - 2g (\ophi^{i}\phi^{i+}
      - \phi^i\ophi^{i-}) \label{D+}\ ,\\
 -D^- &= \sqrt{2}(\pd_i \ophi^{i+} + \opd_i \phi^{i-}) - 2g(\ophi^{i+}\phi^i
        - \phi^{i-}\ophi^{i}) \label{D-}\ .
\end{align}
In a supersymmetric vacuum all auxiliary fields vanish, and vacuum
field configurations can be found by solving a system of first-order
differential equations.


\section{Symmetry breaking and Wilson line}
\label{sec:symmetry}

The condensation of the antisymmetric tensor $\phi^-_0$ breaks $\text{U}(16)$
to $\text{USp}(16)$ \cite{Li:1973mq}. To understand this, it is convenient to consider
the following decomposition of the $\text{SO}(32)$ generators. 
The generators of the unitary subgroup $\text{U}(16)$ are
hermitian matrices $S+iA$ and those of the coset $\text{SO}(32)/\text{U}(16)$ are
complex antisymmetric matrices $A_1 + iA_2$, respectively. Here $S$
denotes real $16 \times 16$ matrices, and $A$, $A_{1,2}$ are antisymmetric
real matrices. They can be expressed as products of Pauli matrices and
$8\times 8$ symmetric ($S'$) and antisymmetric ($A'$) matrices:
\begin{equation}
\begin{split}
S &: \quad \sigma_0 \otimes S'\ , \quad \sigma_1\otimes S' \
    ,\quad i\sigma_2 \otimes A'\ , \quad \sigma_3\otimes S'\ ,\\
A &: \quad \sigma_0 \otimes A'\ , \quad \sigma_1\otimes A' \
    ,\quad i\sigma_2 \otimes S'\ , \quad \sigma_3\otimes A'\ .
  \end{split}
\end{equation}

Tachyon condensation removes the tachyon if the condensate cancels the
flux contribution to the D-term \eqref{D0},
\begin{equation}
  0 =  f \mathbb{1} - g \langle\ophi^-\rangle\langle\phi^-\rangle \ .
\end{equation}
This is achieved by the expectation value
\begin{equation}\label{vevfi-}
  \langle\phi^-\rangle = \sqrt{\frac{f}{g}} \sigma_2\otimes \mathbb{1}'
  \equiv \sqrt{\frac{f}{g}} \hat{\sigma}_2\ .
\end{equation}  
To find the remaining unbroken subgroup one has to determine the
$\text{U}(16)$ generators which commute with this expectation value,
\begin{equation}
  [\langle \hat{\phi}^-\rangle,\hat{V}] =
  \sqrt{2}\left(\begin{array}{c c}
          0 &  0\\ \langle \phi^{-}\rangle V + V^T\langle \phi^-\rangle  & 0
        \end{array}\right) = 0 \ .
\end{equation}      
One easily verifies that $\langle \hat{\phi}^-\rangle$ commutes with $\hat{V}$
for $V=U$, i.e., $\langle \phi^{-}\rangle U + U^T\langle \phi^-\rangle
= 0$, where
\begin{equation}
U : \quad i\sigma_0 \otimes A'\ , \quad \sigma_1\otimes S' \
,\quad \sigma_2 \otimes S'\ , \quad \sigma_3\otimes S'\ .
\end{equation}
These matrices generate the group $\text{USp}(16)$ \cite{Georgi:1982jb}; the related
matrices $\hat{\sigma}_2 U$ are asymmetric. The coset $\text{U}(16)/\text{USp}(16)$ is
generated by $V=X$, where
\begin{equation}
X : \quad \sigma_0 \otimes S'\ , \quad i\sigma_1\otimes A' \
,\quad i\sigma_2 \otimes A'\ , \quad i\sigma_3\otimes A'\ .
\end{equation}
The matrices $\hat{\sigma}_2X$ are antisymmetric, and one has
\begin{equation}\label{Xrelation}
  \langle \phi^- \rangle X = X^T   \langle \phi^- \rangle\ .
\end{equation}

The tachyonic instability will lead to a condensate of $\phi^-$, with
a backreaction on the flux $\phi$, and in general also on
$\phi^+$. Often times, it is argued that after tachyon condensation the
magnetic flux vanishes, and that also supersymmetry is
restored. Hence, one might think that the breaking from
$\text{SO}(32)$ to $\text{USp}(16)$
can be realized by means of the vacuum expectation value (vev)
\begin{equation}\label{Wilsonlinevev}
\langle\hat{\phi} \rangle =
  \left(\begin{array}{c c}
          \phi \, \mathbb{1} & \sqrt{2}\vf_+\hat{\sigma}_2 \\
          \sqrt{2}\hat{\sigma}_2\vf_- & -\phi \, \mathbb{1}
        \end{array}\right) \ ,
\end{equation}  
with $\opd \phi = 0$. For constant vev's $\phi$, $\vf_+$ and $\vf_-$, this
corresponds to the choice of a particular Wilson line.

Supersymmetric vacua are characterized by vanishing F- and
D-terms. Assuming that only one chiral superfield, $\hat{\phi}^3 =
\hat{\phi}$, acquires a vev, all F-terms vanish automatically. The
requirement of vanishing D-terms corresponds to the first-order
differential equations 
\begin{align}
\pd \ophi + \opd \phi
 - \sqrt{2}g( \ovf_{-}\vf_{-}  -  \vf_{+}\ovf_{+}) \label{D01} &=0 \ , \\
(\pd  - \sqrt{2}g \phi)\ovf_+ + (\opd + \sqrt{2}g\ophi)\vf_- &= 0 \label{D-1}\ .
\end{align}
For constant vev's, $\phi = v_0/\sqrt{2}$ and $\vf_\pm = v_\pm/\sqrt{2}$, Eq.~\eqref{D-1}
becomes a constraint on the vev's,
\begin{equation}
  v_0 v^*_+ = v_- v^*_0 \ , \label{vevconstraint}
\end{equation}  
which automatically solves Eq.~\eqref{D01}, $|v_+|^2 = |v_-|^2$.

It is straightforward to compute the masses of fermions, vectors and
scalars in a background of constant vev's. For the quadratic part of
the superpotential one finds, with $\phi^i = \phi^i_U + \phi^i_X$
($i,j=1,2$),
\begin{equation}\label{Wm}
  \begin{split}
  W_{m} \supset -\frac{1}{4}\E_{ij} \int d^2z \, 
  \Tr &\big[\hat{\phi}^i\big(\pd\hat{\phi}^j
    +\frac{g}{\sqrt{2}}[\langle\hat{\phi}\rangle,\hat{\phi}^j]  \big) \big] \\
=-\frac{1}{2}\E_{ij} \int d^2z \,
\tr &\big[\phi^i_U\pd\phi^j_U + \phi^i_X\big(\pd\phi^j_X
+ g(v_+\phi^{j-}-v_-\phi^{k+})\big)\\
&+2\phi^{i+}(\pd - gv_0)\phi^{j-} + g(v_-\phi^{i+}-v_+\phi^{i-})\phi^j_X\big] \ .
\end{split}
\end{equation}
The Kaluza-Klein mode functions are given by 
\begin{equation}\label{KK}
\lambda_{l,m}(z) = e^{zM_{l,m}-\bz\overline{M}_{l,m}} \ , \quad
M_{l,m} =2\pi(m+il) \equiv M_\kk \ .
\end{equation}
 Introducing $\kk = (l,m)$, with $l>0, m \in \mathbb{Z}$ or $l=0, m>0$,
 and $\bk = (-l,-m)$, one has $\lambda_{\bk} = \overline{\lambda}_{\kk}$ and
\begin{equation}
  \pd \lambda_\kk = M_\kk \lambda_\kk \ , \quad
\opd \lambda_\kk = -\overline{M}_\kk \lambda_\kk \ .
\end{equation}
The mode expansion of a chiral superfield $\phi$ takes the form
\begin{equation}
\phi =   \phi_0 + \sum_{\kk}(\phi_{\kk} \lambda_{\kk} + \phi_{\bk}
\lambda_{\bk}) \ . 
\end{equation}
Inserting this expansion into Eq.~\eqref{Wm} one finds
\begin{equation}
W_{m} = - \E_{ij}
\sum_\kk\tr\big[\phi^i_{U\bk}M_\kk\phi^j_{U\kk}
+(\phi^i_{X\bk} ,\phi^{i-}_{\bk} ,\phi^{i+}_{\bk} )\mathcal{M}_f
(\phi^j_{X\kk} ,\phi^{j+}_\kk ,\phi^{j-}_\kk )^T\big] \ ,
\end{equation}
where the fermion mass matrix is given by
\begin{equation}
\mathcal{M}_f =
  \left(\begin{array}{c c c}
          M_\kk & -gv_- & gv_+ \\
          -gv_+ & M_\kk + gv_0 & 0 \\
          gv_- & 0 & M_\kk - gv_0
        \end{array}\right) \ .
\end{equation}  
Its determinant factorizes as 
\begin{equation}
  \text{det} \mathcal{M}_f = M_\kk (M^2_\kk - g^2(v_0^2+2v_+v_-))\ .
\end{equation}
The matrix $\mathcal{M}_f$ describes the mixing of six antisymmetric
tensors $\phi^{i\pm}_\kk$ and $\phi^i_{X\kk}$, $i = 1,2$. One
eigenvalue is $M_\kk$, the other two are $M_\kk \pm g\sqrt{v_0^2 +
  2v_+v_-}$. For $v_\pm =0$ this corresponds to the standard
spectrum of an Abelian Wilson line: for $M_\kk = 0$, there is a pair
of massless chiral multiplets; otherwise, always four chiral
multiplets form degenerate Dirac mass terms. For $v_\pm \neq 0$,
there still is one pair of zero modes for $M_\kk = 0$.
This is surprising, since naively one would expect that the vev's
of the antisymmetric tensors $\phi^\pm$ break $\text{SO}(32)$ to
$\text{USp}(16)$, as discussed above, and all multiplets become massive.
However, the massive spectrum
corresponds to an unbroken group $\text{U}(16)$. We conclude that for any
values  $v_0$, $v_+$ and $v_-$ that satisfy Eq.~\eqref{vevconstraint},
the vacuum expectation value \eqref{Wilsonlinevev} represents an
Abelian Wilson line, just with different embeddings into $\text{SO}(32)$.
The scalar mass matrix is given by
\begin{align}
\mathcal{M}^2_S &= \mathcal{M}^\dg_f \mathcal{M}_f  \nonumber\\
&= \left(\begin{array}{c c c}
  |M_\kk|^2 + 2g^2|v_+|^2 & -g(v_-\overline{M}_\kk +
  v^*_+M_{\kk+}) &  g(v_+\overline{M}_\kk + v^*_-M_{\kk-})\\
          -g(v^*_- M_\kk + v_+\overline{M}_{\kk+})
  & |M_{\kk+}|^2 + g^2|v_+|^2 & -g^2 v_-^*v_+ \\
          g(v^*_+ M_\kk + v_-\overline{M}_{\kk-}) & -g^2 v_+^*v_-
                       & |M_{\kk-}|^2 + g^2|v_+|^2
        \end{array}\right) ,
    \end{align}  
where the relations  $v_0 v^*_+ = v_- v^*_0$ and $|v_+|^2 = |v_-|^2$
have been used, and $M_{\kk\pm} \equiv M_\kk \pm gv_0$.

In order to verify that supersymmetry is unbroken one
also has to evaluate the vector boson masses. Analogous to the
superpotential terms, the starting point is now
\begin{equation}
\begin{split}  
  &\pd \hat{V} + \frac{g}{\sqrt{2}}[\langle\hat{\phi}\rangle,\hat{V}] =\\
  &\left(\begin{array}{ c c}
          \pd U + \pd X + g(v_+X^- - v_-X^+) &
          \sqrt{2}\big((\pd + gv_0)X^+ - gv_+X\big)\hat{\sigma}_2 \\
          \sqrt{2}\hat{\sigma}_2\big((\pd - gv_0)X^- + gv_-X\big)&
           -\big(\pd U + \pd X + g(v_+X^- - v_-X^+)\big)^T
        \end{array}\right).
\end{split}
\end{equation}
Here we have introduced the fields
\begin{equation}
  V^- = \hat{\sigma}_2 X^-\ , \quad V^+ = X^+\hat{\sigma}_2\ , \quad
  X^- = X^{+\dg}\ ,
  \end{equation}
  and we have used the relation \eqref{Xrelation}, 
  $\hat{\sigma}_2 X^T = X \hat{\sigma}_2$. The part of the Kaehler
  potential containing the vector-boson masses is given by
  \begin{equation}\label{Km}
    K_m = \frac{1}{2}\int d^2z \, \Tr\big[
  \big(\pd \hat{V} + \frac{g}{\sqrt{2}}[\langle\hat{\phi}\rangle,\hat{V}]\big)^\dg 
  \big(\pd \hat{V} +
  \frac{g}{\sqrt{2}}[\langle\hat{\phi}\rangle,\hat{V}]\big)\big]\ . 
\end{equation}
In the KK mode expansion (see Eq.~\eqref{KK}) one has to make sure that the relations
$X^\dg = X$ and $X^{+\dg} = X^-$ are satisfied, which yields
\begin{equation}
\begin{split}
X &= X_0 + \sum_{\kk} (X_{\kk}\lambda_{\kk} + X_{\bk} \lambda_{\bk})
\ , \quad X_{\bk} = X^\dg_{\kk} \ , \\
X^\pm &= X^\pm_0 + \sum_{\kk} (X^\pm_{\kk}\lambda_{\kk}
+ X^\pm_{\bk} \lambda_{\bk})
\ , \quad X^{\pm\dg}_{\kk} = X^\mp_{\bk} \ . 
\end{split}
\end{equation}
Inserting this expansion into Eq.~\eqref{Km}
one finds
\begin{equation}
K_m = 2 \sum_{\kk}\tr\big[|M_\kk|^2U_{\bk} U_{\kk} + (X_{\bk} , X^-_{\bk} , X^+_{\bk} )
\mathcal{M}^2_V (X_{\kk} , X^+_{\kk} , X^-_{\kk} )\big] \ ,
\end{equation}
where the vector-boson mass matrix indeed satisfies
\begin{equation}
  \mathcal{M}^2_V = \mathcal{M}^\dg_f \mathcal{M}_f = \mathcal{M}^2_S
  \ .
\end{equation}
A pair of massive 4d $\mathcal{N}=1$ vector multiplets is mass
degenerate with four 4d $\mathcal{N}=1$ chiral multiplets,
corresponding altogether to eight Weyl fermions, six complex
scalars and a complex vector. These degrees of freedom form
a massive 8d $\mathcal{N}=1$ multiplet. The massless multiplet
contains half the number of degrees of freedom. We
  conclude that Wilson lines can break $\text{SO}(32)$ only to
  $\text{U}(16)$, but not to $\text{USp}(16)$.


\section{Dilute vs. singular flux}
\label{sec:singular}

In the magnetic flux background discussed in
Section~\ref{sec:magnetic}, only the fields $\phi^-_{0,j}$ are
tachyonic where $j=0,\ldots,2k-1$ labels the degeneracy for $k$ flux quanta.
This suggests to search for solutions of the equations of motion where only
$\phi^-_{0,j}$ acquire a non-zero vev. Searching for solutions with unbroken
supersymmetry, we again consider the first-order differential equations
corresponding to vanishing D-terms, which now read
($\langle\phi^-_{0,j}\rangle = \hat{\sigma}_2\vf_{-,j}$, $\vf_{-,j}  \equiv
\vf_j$)
\begin{align}
\pd \ophi + \opd \phi
 - \sqrt{2}g \sum_{j=0}^{2k-1}\ovf_j\vf_j  =0 \ , \quad 
(\opd + \sqrt{2}g\ophi)\vf_j = 0 \label{Ds}\ .
\end{align}

First, we consider these equations on the covering space $\mathbb{C}$.
The equation for $\vf_j$ can be solved with the
ansatz\footnote{Such an ansatz has previously been considered for
  zero-mode wave functions on a magnetized orbifold \cite{Lee:2003mc, Buchmuller:2018lkz}.}
\begin{equation}
  \phi = \frac{1}{\sqrt{2}}\pd \chi\ , \quad \vf_0 = \vf = \beta z^b e^{-g\chi} \ , \quad \vf_{j \geq1} = 0 \ ,  \quad \chi = \bar{\chi}\ ,
\end{equation}
which leads to a second-order differential equation for $\chi$,
\begin{equation}
  \pd\opd \chi = g \beta^2 (\bar{z}z)^b e^{-2g\chi}\ .
\end{equation}  
After introducing polar coordinates, $\pd\opd =
r^{-1}\pd_r(r\pd_r) + r^{-2}\pd^2_\theta$, changing $r$ to $t=\ln{r}$,
and defining $h = \chi - (b+1)t/g$, this becomes
\begin{equation}\label{diffh}
  \pd_t^2 h(t) = g \beta^2  e^{-2g h(t)}\ .
\end{equation}
A straightforward calculation yields the result
\begin{equation}
gh =
\ln{\left(\frac{\beta}{2^{b+1}C}\left(\left(\frac{|z|}{|z_0|}\right)^{gC}
+\left(\frac{|z_0|}{|z|}\right)^{gC}\right)\right)}\ ,
\end{equation}
where $C>0$ and $|z_0|$ are integration constants. This yields for
$g\chi = gh +(b+1)t$,
\begin{equation}
g\chi =
\ln{\left(\frac{\beta |z_0|^{b+1}}{C}\left(
      \left(\frac{|z|}{|z_0|}\right)^{gC+b+1}+\left(\frac{|z_0|}{|z|}\right)^{gC-b-1}
    \right)\right)}\ .
\end{equation}
Demanding that the solution is regular at $z=0$ fixes the integration
constant $C$ as $C=(b+1)/g$. 
One then obtains for the fields $\phi$ and $\vf$
\begin{equation}\label{fullsolution}
\begin{split}
\phi &=\ \frac{b+1}{\sqrt{2}gz} \frac{|z|^{2(b+1)}}
{|z_0|^{2(b+1)}  + |z|^{2(b+1)}} \ ,\\
\vf &=\ \frac{b+1}{g} z^b \frac{|z_0|^{b+1}}
{|z_0|^{2(b+1)}  + |z|^{2(b+1)}} \ .
\end{split}
\end{equation}

One might think that the  flux could somehow disappear in the process of tachyon condensation. This would be the case if, by varying continuously a parameter, the solution interpolates
between the magnetic field and the usual vacuum, with no magnetic flux and no tachyon condensation. This is however not the case. For general integration constant $C$ the classical solution
is singular, so $C$ cannot vary continuously.  The trivial vacuum solution, with no magnetic flux, is of course also a solution of the D-term
equations (\ref{Ds}). But it is an isolated solution, not connected by any parameter deformation to the classical solution (\ref{fullsolution}).

The constant $z_0$ defines a length scale that separates the complex
plane into an inner core, $|z| < |z_0|$ and an outer region, $|z| >
|z_0|$. For $b=0$, one obtains in the inner core
$|z| \ll |z_0|$ a constant flux density and
constant expectation value $\vf$ (cf.~Eqs.~\eqref{vev}, \eqref{vevfi-}),
\begin{equation}
  \phi = \frac{f\bz}{\sqrt{2}}\Big(1 + \mathcal{O}\Big(\frac{|z|^2}{|z_0|^2}\Big)
\Big)\ , \quad
  \vf = \sqrt{\frac{f}{g}} \Big(1 + \mathcal{O}\Big(\frac{|z|^2}{|z_0|^2}\Big)
\Big) \ ,
  \end{equation}
where $f = 1/(g|z_0|^2)$. Note that for this perturbative solution in
powers of $|z|^2/|z_0|^2$ the D-term vanishes.

The radius $|z_0|$ plays the role of a regulator and the limit $|z_0|
\rightarrow 0$ is of particular interest. In this limit $\phi$ and
$\vf$ are singular and from Eq.~\eqref{fullsolution} one obtains
\begin{equation}
  \phi \rightarrow \frac{b+1}{\sqrt{2}gz}\ ,\quad
  \vf^2 \rightarrow \frac{4(b+1)\pi}{g^2} \delta^{(2)}(z,\bar{z})\ .
\end{equation}
$\phi = (A_9+iA_8)/\sqrt{2}$ describes the vortex vector potential,
\begin{equation}\label{singularvector}
 A_m = -\frac{2(b+1)}{g}\epsilon_{mn}\frac{x_n}{x_8^2+x_9^2}  .
\end{equation}
The flux vanishes everywhere except at the origin, and the localized
flux density is given by a line integral over a 1-cycle $\mathcal{C}_O$ around
the origin\footnote{Since the charge of the antisymmetric
    tensor field is $q=2$, the factor $4\pi$ occurs instead of $2\pi$.}, 
\begin{equation}
g\oint_{\mathcal{C}_O} A = 4 \pi (b+1) \ .
\end{equation}
On the plane the flux is not quantized and therefore $b$ can take
arbitrary values. On the torus flux
quantization requires $b+1 = k/2$, $k\in \mathbb{Z}$, which will be discussed in
the next section. Note that a singular wave function\footnote{We use
  the notation $\delta^{(2)}(z,\bar{z}) \equiv \delta(x_8, x_9)$.} $\vf \propto
\delta^{1/2}$ has previously been considered
in connection with the infinite flux limit on a magnetized
torus \cite{Cremades:2004wa}.

Notice that for $b \not=0$ the classical solution has no
  rotational symmetry around the core $z=0$.  Also, while in the zero-size limit $z_0 \to 0$ the solution has the same localized
flux interpretation for any $b$, in the opposite limit $z_0 \gg z$ it does not correspond anymore to a constant magnetic field unless $b=0$. Based on the string theory
intuition in Section 6  and on the analogy with instanton solutions, we believe that for a general magnetic flux, the interpolation between a constant magnetic field and  the localized flux should
start from a multicenter solution, with a different ansatz. We were
unable to find explicitly the solution but hope to come back to this point in the future. 

It is often assumed that tachyon condensation corresponds to the
annihilation of two branes such that the total magnetic flux
vanishes. Our result is different. We find that the total flux of the
unstable state is still present after tachyon condensation, but it is
squeezed into a singularity. As we shall see below, this affects the
boundary conditions of fields in this vacuum and therefore the massive spectrum.

Finally, since this classical solution was obtained by ignoring the gravitational sector, one might worry about the gravitational backreaction that could change it significantly. However, the Yang-Mills classical
solution is supersymmetric and its energy is zero. Therefore the gravitational fields will not be excited and the spacetime geometry remains flat. 

\begin{figure}[t]
\begin{center} 
\includegraphics[width = 0.4\textwidth]{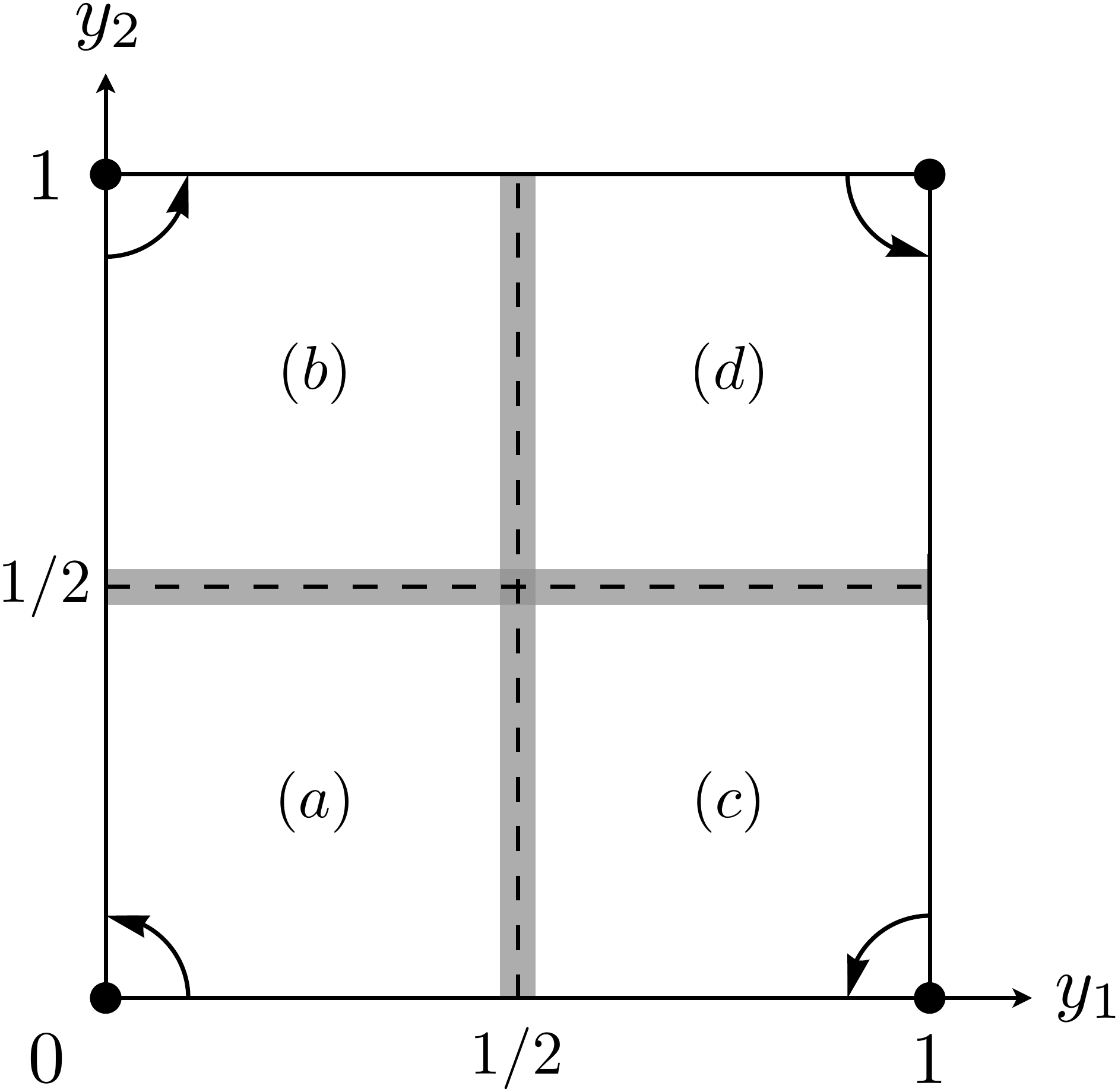}
\end{center}
\caption{Vortex vector potential with singularity at $(y_1,y_2) = (0,0)$ 
on a square torus covered by four patches.}
\label{fig:singular}
\end{figure}

\section{Massive spectrum for singular flux}
\label{sec:massive}

In the previous section we have constructed a solution of the equations
of motion for $\phi$ and $\phi^-$ in the complex plane. Let us now
consider the singular limit of this solution as background
fields on the torus. The gauge background $\phi$ corresponds to the
vortex vector potential depicted in Fig.~\ref{fig:singular}. The field
$\phi^-$ is localized at the origin.

The vector potential can be expressed in terms of the singular
function\footnote{For convenience, we now rescale $A_{8,9} \rightarrow
  A_{1,2}/(2\pi  R)$, $x_{8,9} \rightarrow 2\pi Ry_{1,2}$.}
\begin{equation}\label{singlambda}
  \Lambda(y_1,y_2) = i\frac{k}{2g} \ln\left(\frac{y_1 +
        iy_2}{y_1-iy_2}\right) = \overline{\Lambda}(y_1,y_2) \ ,
\end{equation}  
where $k = 2(b+1)$ is determined by the flux density (see Eq.~\eqref{singularvector})
In the four patches one has (see Eq.~\eqref{singularvector})
\begin{equation}\label{transition}
\begin{split}
(a)\quad A^a_m(y_1,y_2) &=-\partial_m \Lambda_a(y_1,y_2)\ , \quad    
\Lambda_a(y_1,y_2) = \Lambda(y_1,y_2) \ ,\\
(b)\quad A^b_m(y_1,y_2) &=-\partial_m \Lambda_b(y_1,y_2)\ , \quad    
\Lambda_b(y_1,y_2) = \Lambda(y_1,y_2-1) \ , \\
(c)\quad A^c_m(y_1,y_2) &=-\partial_m \Lambda_c(y_1,y_2)\ , \quad    
\Lambda_c(y_1,y_2) = \Lambda(y_1-1,y_2) \ ,\\
(d)\quad A^d_m(y_1,y_2) &=-\partial_m \Lambda_d(y_1,y_2)\ , \quad    
\Lambda_d(y_1,y_2) = \Lambda(y_1-1,y_2-1) \ .
\end{split}
\end{equation}
The transition functions are easily obtained from
Eqs.~\eqref{singlambda}, \eqref{transition} and in Appendix A \eqref{overlap}. For
example, for the patches $(a)$ and $(b)$ one has
\begin{equation}
(\partial_m \Lambda_{ba})(y_1,1/2) = g (A^a_m(y_1,1/2) -  A^b_m(y_1,1/2))
= - g (\partial_m\Lambda_{a}-\partial_m\Lambda_{b})(y_1,1/2)\ ,
\end{equation}
which yields, up to a constant,
\begin{equation}
\begin{split}  
\Lambda_{ba}(y_1,1/2) &= -g (\Lambda_{a}(y_1,1/2)
-\Lambda_{b}(y_1,1/2)) \\
&= ik \ln\left(\frac{y_1-i/2}{y_1+i/2}\right) = 2k \arctan(1/(2y_1)) \ .
\end{split}
\end{equation}
The charged fields in the patches $(a)$ and $(b)$ are related by the
transition function,
\begin{equation}\label{phiminusab}
  \phi^-_b (y_1,1/2) = S^{-1}_{ba}(y_1,1/2)  \phi^-_a (y_1,1/2) =
  e^{-i\Lambda_{ba}(y_1,1/2)}  \phi^-_a (y_1,1/2)\ .
\end{equation}
For the singular solution the charged field is localized\footnote{We
  use the notation $\delta^{(2)}(z,\bar{z}) \equiv \delta(y_1,y_2)$.},
  $\phi^-_a(y_1,y_2) \propto \delta^{1/2}(y_1,y_2)$.
 Analogously, in patch $(b)$ one has
 $\phi^-_b(y_1,y_2) \propto \delta^{1/2}(y_1,y_2-1)$. Hence, in
 the singular limit the field $\phi^-$
 vanishes in the bulk and the condition \eqref{phiminusab} is
 trivially satisfied. The same holds for all boundaries between
 different patches, which means that the singular background fields
 represent a solution of the equations of motion on the torus.
 The consistency condition at the overlap of the four patches,
\begin{equation}
S_{ac}(1/2,1/2)S_{cd}(1/2,1/2)
S_{db}(1/2,1/2)S_{ba}(1/2,1/2)
= e^{-i2\pi k} =1\ ,
\end{equation}
implies $k\in \mathbb{Z}$ and therefore standard flux quantization
also for the localized flux.

In order to obtain the massive spectrum of the 8d theory one has to
shift in the 10d Lagrangian \eqref{L10d} the fields $\phi$ and
$\phi^-$ around the background fields $\langle \phi\rangle$ and
$\langle \phi^-\rangle$, respectively, and keep the quadratic part of
the Lagrangian, which reads
\begin{equation}\label{LVm}
\begin{split}
  \mathcal{L}_{m} \supset
  \int d^4 \theta \, \tr \big[
&\big(\opd V - \sqrt{2}g \langle\ophi^-\rangle V^- \big) 
\big(\pd V - \sqrt{2}g V^+\langle\phi^-\rangle \big)\\
&+ \big(\opd V^+- \sqrt{2}g\langle\ophi\rangle V^+ +\frac{g}{\sqrt{2}}(
   V\langle\ophi^{-}\rangle +\langle\ophi^{-}\rangle V^T)\big)\\
&\quad\times
\big(\pd V^-- \sqrt{2}g \langle\phi\rangle V^-
+ \frac{g}{\sqrt{2}}(\langle\phi^{-}\rangle V+V^T\langle\phi^{-}\rangle)\big) \\
&+\big(\opd V^- + \sqrt{2}g\langle\ophi\rangle V^-\big)
\big(\pd V^+ +\sqrt{2}g\langle\phi\rangle V^+\big)\big] \ .
\end{split}
\end{equation}
With $V = U + X$, where $U \in \text{USp}(16)$ and $X \in \text{U}(16)/\text{USp}(16)$, one has
\begin{equation}
V\langle\ophi^{-}\rangle +\langle\ophi^{-}\rangle V^T =
2X \langle\ophi^{-}\rangle\ , \quad
\tr [U V^+\langle \phi^-\rangle] = 0\ .
\end{equation}
Therefore $U$ does not mix with charged fields, and defining
$\langle\ophi^{-}\rangle = \hat{\sigma}_2\vf$, 
$V^- = X^- \hat{\sigma}_2$ and $V^+ = \hat{\sigma}_2 X^+$, one obtains
\begin{equation}\label{LVm2}
\begin{split}
  \mathcal{L}_{m} \supset
  \int d^4 \theta \, \tr \big[
  & \opd U \pd U + 
  \big(\opd X - \sqrt{2}g \vf X^- \big) \big(\pd X - \sqrt{2}g \bar{\vf}X^+\big)\\
&+ \big((\opd -\sqrt{2}\langle \bar{\phi}\rangle)X^++ \sqrt{2}g\vf X\big)
\big((\pd - \sqrt{2} g \langle\phi\rangle) X^-+ \sqrt{2}g \bar{\vf} X\big)\\
&+ (\opd + \sqrt{2} g \langle\bar{\phi}\rangle )X^-
(\pd + \sqrt{2}g\langle\phi\rangle) X^+\big] \ .
\end{split}
\end{equation}
In the background gauge field $\langle\phi\rangle$, the charged fields $X^\pm$ are 
also defined in the four patches. At the overlaps they are related by
gauge transformations, for instance,
\begin{equation}
  X^+_b(y_1,1/2) = S_{ba}(y_1,1/2) X^+_a(y_1,1/2)
  = e^{i\Lambda_{ba}(y_1,1/2)} X^+_a(y_1,1/2) \ .
\end{equation}  

The singular vector field defined in Eq.~\eqref{transition} can be
removed by a singular redefinition of the charged fields
\begin{equation}
   X^+_a(y_1,y_2) = e^{ig\Lambda_a(y_1,y_2)}  \hat{X}^+_a(y_1,y_2)\ , \quad
 X^+_b(y_1,y_2) = e^{ig\Lambda_b(y_1,y_2)}  \hat{X}^+_b(y_1,y_2)\ , \ldots
\end{equation}
The new fields have trivial transition functions\footnote{This is
  analogous to the transition from `twisted' wave functions to
  `untwisted' wave functions on a magnetized torus, as discussed in
  Ref.~\cite{Buchmuller:2018lkz}.} , e.g.,
\begin{equation}
  \begin{split}
  \hat{X}^+_b(y_1,1/2) &= e^{-ig\Lambda_b(y_1,1/2)} X^+_b(y_1,1/2)
  = e^{-ig\Lambda_b(y_1,1/2)+i\Lambda_{ba}(y_1,1/2)} X^+_a(y_1,1/2)\\
&= e^{-ig\Lambda_a(y_1,1/2)} X^+_a(y_1,1/2) = \hat{X}^+_a(y_1,1/2) \ .
\end{split}
\end{equation}  
In terms of the fields $\hat{X}^\pm$ the Lagrangian \eqref{LVm2} becomes
\begin{equation}\label{LVm3}
\begin{split}
  \mathcal{L}_{m} \supset
  \int d^4 \theta \, \tr \big[
  & \opd U \pd U + 
  \big(\opd X - \sqrt{2}g \vf' \hat{X}^- \big)
  \big(\pd X - \sqrt{2}g \bar{\vf}'\hat{X}^+\big)\\
&+ \big(\opd \hat{X}^++ \sqrt{2}g\vf' X\big)
\big(\pd \hat{X}^-+ \sqrt{2}g \bar{\vf}' X\big) + \opd \hat{X}^- \pd \hat{X}^+\big] \ ,
\end{split}
\end{equation}
where $\vf' = \vf \exp{(i\Lambda_O)}$ is the redefined charged background field
that is singular at the origin $O$.

Except for the singular term at the origin, this Lagrangian describes free fields.  
Due to the singular term, the equations of motion require that the
fields $X$ and $\hat{X}^\pm$ vanish at the origin. A basis of periodic $(P)$
and anti-periodic $(A)$ functions on the interval $[0,1]$ is given by
($m \in \mathbb{Z}$),
\begin{equation}\label{PA}
\begin{split}  
(P)\quad c(x) &= \cos{(2\pi mx)}\ , \quad  s(x) = \sin{(2\pi mx)} \ ,\\
(A)\quad \hat{c}(x) &= \cos{(2\pi (m + \tfrac{1}{2}) x)}\ , \quad
\hat{s}(x) = \sin{(2\pi (m + \tfrac{1}{2}) x)} \ .
\end{split}
\end{equation}
From combinations of these periodic and anti-periodic functions one
can form various products that vanish at the origin of the torus. On
the other hand, $U$ has a zero mode and its wavefunctions need not vanish at the origin. One natural way to achieve this is to assign $U$ an even
parity under reflection of torus coordinates, whereas $X, \hat{X}^{\pm}$ are assigned odd parity. An
example of mode functions for the four real fields $U$, $X$,
$X_1 = (\hat{X}^++\hat{X}^-)/\sqrt{2}$ and $X_2 = i(\hat{X}^+-\hat{X}^-)/\sqrt{2}$,
    exhausting all possibilities, is given by
\begin{equation}\label{example}
\begin{split}      
U:\quad &c(y_1)c(y_2)\ ,  \quad s(y_1)s(y_2) \ , \\ 
X_1:\quad &\hat{c}(y_1)s(y_2)\ , \quad \hat{s}(y_1)c(y_2) \   , \\ 
X_2:\quad &c(y_1)\hat{s}(y_2)\ , \quad s(y_1)\hat{c}(y_2)\  ,  \\ 
X:\quad &\hat{c}(y_1)\hat{s}(y_2)\ , \quad \hat{s}(y_1)\hat{c}(y_2) \  . 
\end{split}
\end{equation}
Note that $U$ is periodic, whereas $X_1$, $X_2$ and $X$ satisfy twisted
boundary conditions in $y_1$-direction, $y_2$-direction and both
directions, respectively. The mass spectra $M^2_{m_1,m_2} R^2$ in the
four cases are given by
\begin{equation}\label{massesFT}
\begin{split}
  U:\;\;  & m_1^2+ m_2^2 \ ,   \quad
  X:\;\;  (m_1+\tfrac{1}{2})^2+ (m_2+\tfrac{1}{2})^2 \ ,\\
X_1:\;\;  & (m_1+\tfrac{1}{2})^2 + m_2^2  \ ,
\quad X_2:\;\;  m_1^2 +  (m_2+\tfrac{1}{2})^2  \ .
\end{split}
\end{equation}
The boundary conditions \eqref{example} only represent an example. To
determine them uniquely, information beyond the Lagrangian \eqref{LVm3}
is required, involving interaction terms between the fields and
possibly also couplings to string excitations.


\section{An alternative D9-D7 brane description}
\label{sec:d9d7}
In Section~\ref{sec:singular} we noticed that the classical solution with tachyon condensation has a singular  limit when the length of the core supporting the solution shrinks to zero size. The resulting singular solution
 has a delta-function like flux, which in string theory has the
 natural interpretation of a D7-brane. In the following we therefore
 recall the connection between a magnetized D9-brane and a system of
 elementary D9-D7 branes in type I string theory.

In type IIB, a D9-brane with a magnetic field in its worldvolume
generates a D7-brane charge from the Wess-Zumino coupling
\begin{equation}
\int C \wedge e^F = \int C_{10} + \int C_8 \wedge F  \ , \label{sm2}
\end{equation}
where $C_{p+1}$ denote the RR $(p+1)$-form fields and
  $F$ is the magnetic 2-form field.
In  type I strings  $C_8$ is removed by the orientifold projection $\Omega$. This implies that the BPS D7-brane of type IIB string becomes the non-BPS (uncharged) D7-brane of type I, see \cite{Dudas:2001wd}
for more details of its construction. So let us start from the
corresponding system of
D9- and non-BPS D7-branes in type I.  The gauge group for the  D9-branes and a number $d$ of D7-branes is $\text{SO}(N)_9 \times \text{U}(d)_7$, with $N=32$ due to the RR tadpole condition in type I. Whereas $d$
could in principle be arbitrary since the non-BPS D7-branes have no RR charge, only stacks with $d\leq 16$ are related to the magnetized D9-brane picture, as will be explained below. The low-energy massless 
and tachyonic string spectrum, ignoring massive string excitations, comprises \cite{Dudas:2001wd}
\begin{itemize}
\item supersymmetric gauge multiplets  $\big({\bf
    \frac{N(N-1)}{2},1}\big) + ({\bf 1, d \bar d})$ \ ,
\item chiral fermions  $\big({\bf 1, \frac{d(d-1)}{2}}\big)_R +
  \big({\bf 1, \frac{d(d+1)}{2}}\big)_L + ({\bf N,  d})_R$ \ ,
\item complex scalar tachyons  $\big({\bf 1,
    \frac{d(d-1)}{2}}\big)+ ({\bf N,d})$\ \ .
\end{itemize}
The spectrum can be found in a systematic way from the partition
functions \cite{Dudas:2001wd}, with the light-cone $\text{SO}(8)$
characters decomposed appropriately into products of $\text{SO}(6)$ 
and $\text{SO}(2)$ characters. Using notations and
  conventions of \cite{Angelantonj:2002ct}, the corresponding amplitudes are given by 
\begin{equation}\label{sm02}
  \begin{split}
{\cal A}_{99} &= \frac{N^2}{2} \int_0^{\infty} \frac{d
  \tau_2}{\tau_2^5} \frac{1}{\eta^8}(
  V_6 O_2+ O_6 V_2-S_6 S_2 - C_6 C_2) P_{m_1} P_{m_2} \ ,   \\
{\cal A}_{77} &=  \int_0^{\infty} \frac{d \tau_2}{\tau_2^5}  \frac{1}{\eta^8} (d
{\bar d}(V_6 O_2+ O_6 V_2-S_6 S_2 - C_6 C_2) \\
&\hspace{2.5cm}+\frac{d^2+ {\bar d}^2}{2} (O_6 O_2+ V_6 V_2-S_6 C_2 - C_6 S_2)) 
W_{n_1} W_{n_2}  \ ,   \\
 {\cal A}_{79} &=  \int_0^{\infty} \frac{d \tau_2}{\tau_2^5} \frac{1}{\theta_4\eta^5} ( N d
 (O_6 S_2+ V_6 C_2-C_6 O_2 - S_6 V_2)\\
&\hspace{2.5cm}+N {\bar d} (O_6 C_2+ V_6 S_2-S_6 O_2 - C_6 V_2)) \ ,  \\
{\cal M}_9 &= - \frac{N}{2} \int_0^{\infty} \frac{d \tau_2}{\tau_2^5}
\frac{1}{{\hat \eta}^8} ( {\hat V}_6   {\hat O}_2  + {\hat O}_6 {\hat V}_2  - {\hat S}_6  {\hat S}_2  - {\hat C}_6  {\hat C}_2  )   P_{m_1} P_{m_2} 
\ ,   \\
{\cal M}_7 &= - \frac{1}{2} \int_0^{\infty} \frac{d \tau_2}{\tau_2^5}
\frac{2}{ {\hat \theta}_2 {\hat \eta}^5} ( (d + {\bar d})( {\hat O}_6
{\hat O}_2  + {\hat V}_6 {\hat V}_2 ) \\
&\hspace{3.5cm} + (d - {\bar d})  ({\hat S}_6 {\hat C}_2  - {\hat C}_6 {\hat S}_2))
 W_{n_1} W_{n_2} \ ,
\end{split}
\end{equation}
where $P_m$ and $W_n$ are the momentum and winding sums, respectively.

The connection between the D9-D7 system and the magnetized D9-brane in
type I can be understood in the following way. Start with type I with
an equal number of D9- and D7-branes,
with gauge group $\text{SO}(32)_9 \times \text{U}(16)_7$. By one
T-duality,   one gets the type I' string with 16 BPS D8-branes, with
wrapping numbers $(1,0)$, and 16 non-BPS D8-branes with wrapping numbers $(0,1)$.  
The condensation of the bi-fundamental tachyon $({\bf N,d})$ breaks
the gauge group $\text{SO}(32)_9 \times \text{U}(16)_7$ to the diagonal $\text{U}(16)$.  This can be understood as
a process of recombining  the two
D8-branes, with the recombined brane defined by a cycle with wrapping
numbers $(1,1)$. At this  step, this setup has the interpretation of
intersecting D8-branes in type IIA in eight dimensions, which T-dualized 
back to type I  correspond to a magnetized stack of 16 D9-branes,
of minimum magnetic flux $k=1$ (see left part of Fig.~\ref{fig:USP}).

Similarly, the more general case of magnetized D9-branes in type I with an arbitrary flux integer $k$  has the following description. One starts in type I with
D9-branes and $k$ stacks of $M$ coincident non-BPS D7-branes, $k \leq 16$, at different locations on the 2-torus. The initial gauge group is therefore $\text{SO}(32) \times \text{U}(M)^k$. In the large torus volume limit, there are tachyons in the 
bi-fundamental representation of $\text{SO}(32)$ with all the D7 stacks, but the scalars stretched between different stacks of D7-branes are massive. After one T-duality, one obtains  the type I' string with 16 BPS D8-branes, 
with wrapping numbers $(1,0)$ and $k$ parallel stacks of
$M$ non-BPS $\text{D}8'_i$-branes with wrapping numbers $(0,1)$.  The condensation of the bi-fundamental tachyons stretched between the D8-
and $\text{D}8'_i$-branes break the gauge group $\text{SO}(32)_9 \times
\text{U}(M)^k$ to  $\text{SO}(32-2M) \times \text{U}(M)$.  In this case the recombined branes of gauge group $\text{U}(M)$ have the wrapping numbers $(1,k)$. 
After T-dualizing back to type I, one obtains $16-M$ standard (unmagnetized) D9-branes and a magnetized stack of $M$ D9-branes, with magnetic flux $k$. 
The case of interest for this paper is when all D9-branes are magnetized, i.e.~$M=16$. 
 
The situation is conceptually similar to the D9-D5 or the D5-D1 brane systems first  worked out in \cite{Witten:1995gx,Douglas:1995bn}.  The D9-D5 system for example  has three different but related descriptions. 
In one of them, the $k$ D5-branes are seen as configurations of instanton number $k$ on the D9-brane gauge theory, in the limit where the instanton size goes to zero. In $\text{SO}(32)$ type I strings, 
the D5 gauge group is generically $\text{USp}(2k)$, and the moduli space  coincides precisely with the one of the gauge theory instanton configurations in the $\text{SO}(32)$ D9-brane gauge group. 
The instanton size is a modulus in the supersymmetric gauge theory, which can therefore be varied with no cost. 
For a finite instanton size, the D5-brane is not dynamical anymore. In the maximal instanton-size limit, the physics is captured by a smooth constant magnetic field configuration in the four dimensions transverse to the
D5-brane, of self-dual instanton type $B_1 = B_2$, where $B_{1,2} = k_{1,2}/v_i$ are the constant magnetic fields in the two tori of area $v_i$. 

In type I string theory, a D9-brane with magnetic fields in its worldvolume generates a D5-brane charge from the Wess-Zumino coupling
\begin{equation}
\int C \wedge e^F = \int C_{10} + \int C_6 \wedge F_1 \wedge F_2  \ ,  \label{sm2}
\end{equation}
where $F_{1,2} = B_{1,2} Q$, $Q$ being an Abelian generator of $\text{SO}(32)$. The induced D5 charge in this case is $k=k_1 k_2$.  It is possible to write down explicitly the classical solution of the D9-D5 system
and check the interpolation between the zero-size instanton limit, with D5-branes corresponding to a singular source in the limit, and the maximum size limit, where the solution describes self-dual constant magnetic
fields in the two tori \cite{Angelantonj:2011hs}.  

The difference between the D9-D5 and the D9-D7 systems is that in the first case the instanton size, equivalent to the D9-D5 hypermultiplet vev's, is a flat direction due to supersymmetry and  the limit of small 
vev can be captured by an effective field theory description. For the D9-D7 system, due to the stringy value of the  tachyonic mass, the process cannot be captured reliably by field theory. 
However, as we saw in Section~\ref{sec:singular},  for the magnetized
D9-brane description on the torus with minimum magnetic quantum number, there is a similar classical solution parametrized by a parameter $z_0$, 
which is similar to the instanton size of the D9-D5 brane system. Indeed, in the limit $z_0 \to 0$, the solution describes a localized source and a localized vev for the negatively charged fields at the origin of the torus, 
which are the D7-branes. On the other hand,  in the large  $z_0$ limit, the solution becomes a constant magnetic field on the D9-branes, plus a profile for the tachyon which approaches a constant in 
the $z_0 \gg R$ limit. This fits with the tachyon condensation in the D9-D7 system above. 
In Section 4, we constructed a classical solution for
  arbitrary flux that has a singular limit for $z_0 \to 0$
  corresponding to a localized flux, but we do not know the classical solution interpolating from this singular solution to a constant magnetic field. As we mentioned in Section 4,  such a solution could correspond to  several localized sources in the singular
limit.

Finally, tachyon condensation in the antisymmetric
representation
$ (\bf 1, d(d-1)/2)$ breaks $\text{U}(16)$ to $\text{USp}(16)$.  The final state is supersymmetric, so a natural question is the string
theory interpretation of it, if any. Fortunately, there is only one
candidate superstring in 8d, which does have gauge group $\text{USp}(16)$, to which we now turn.

\section{The 8d USp(16) superstring as magnetized\\ type I superstring
  after tachyon condensation}
\label{sec:unique}

\begin{figure}[t]
\begin{center} 
\includegraphics[width = 0.8\textwidth]{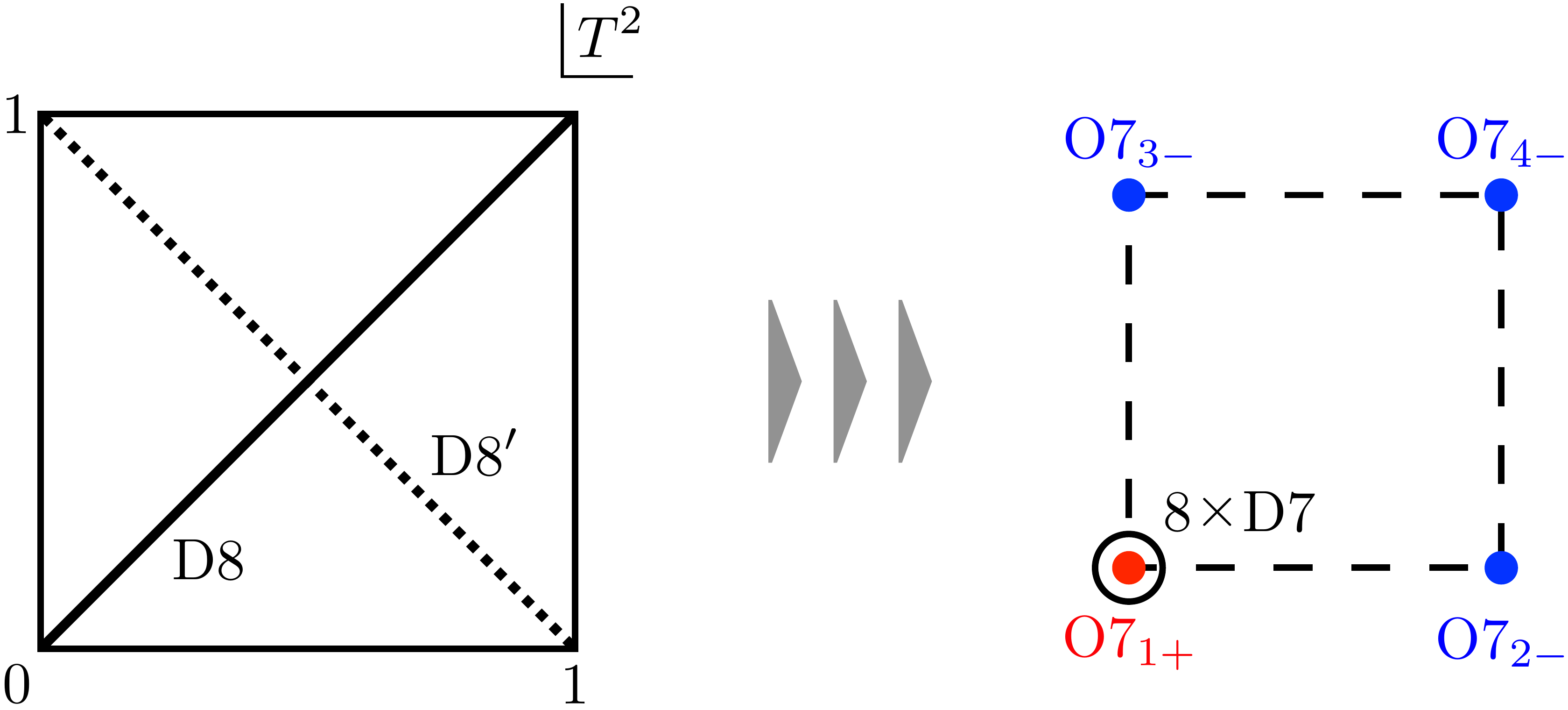}
\end{center}
\caption{The left picture represents the type IIA orientifold with
  intersecting D8-branes. After tachyon condensation and one T-duality
  the conjectured final theory on the right is a type IIB orientifold
  with D7- and O7-planes where tachyon condensation changes one
  $\text{O}7_-$-plane into an $\text{O}7_+$-plane. }
\label{fig:USP}
\end{figure}

To our knowledge there exists only one eight-dimensional superstring theory that can be the ground state after tachyon condensation.  Its gauge group is $\text{USp}(16)$, the group obtained after
tachyon condensation as described in Section~\ref{sec:singular}. This
8d string theory was first constructed by Bianchi, Pradisi and
Sagnotti \cite{Bianchi:1991eu,Bianchi:1997rf}, and its geometry was understood later on by Witten using a T-dual picture \cite{Witten:1997bs}. 
The theory has also a dual description in terms of CHL strings \cite{Chaudhuri:1995fk}.
The original description \cite{Bianchi:1991eu,Bianchi:1997rf}  makes
use of the antisymmetric tensor field $B_{ab}$.  This field is odd
under world-sheet parity  and therefore it is projected out by the
orientifold projection $\Omega$ in type I superstring. However, this still leaves
the possibility to add a quantized value
$\frac{2}{\alpha'} B_{ab} \in \mathbb{Z}$. The corresponding 8d theory is supersymmetric and, in the
absence of additional Wilson lines, it has gauge group
$\text{USp}(16)$ from the open string sector.

The T-dual version of the resulting theory has a
particularly simple interpretation \cite{Witten:1997bs}. In the
presence of the discrete background  $B_{ab}$ T-duality in two coordinates
acts as $R_1 \to \frac{\alpha'}{2 R_1}$, $R_2 \to \frac{\alpha'}{2 R_2}$.
The T-dual of the standard $\text{SO}(32)$ type I
superstring turns the original $\text{O}9_-$-plane wrapping the torus into four
$\text{O}7_-$-planes at orientifold fixed points. On the contrary, the T-dual of the
$\text{USp}(16)$ superstring  turns the original $\text{O}9_-$-plane into
three $\text{O}7_-$-planes and one $\text{O}7_+$-plane on the orientifold. While an
$\text{O}7_-$-plane has charge (and tension) equal to $-4$ in units of a
regular D7-brane charge,  an $\text{O}7_+$-plane has charge (and tension) 
equal to $+4$.  The switch $\text{O}7_- \to \text{O}7_+$ has the
overall effect of halving the RR tadpole and therefore the rank of the
gauge group, a fact that requires addition of only eight D7-branes.  
Furthermore, while  D7-branes on top of an $\text{O}7_-$-plane lead to an
orthogonal (SO) gauge group, D7-branes on top of an $\text{O}7_+$-plane
lead to a symplectic (USp) gauge group. 
Therefore, the configuration with all the D7-branes sitting on top of
the $\text{O}7_+$-plane leads to the gauge group $\text{USp}(16)$. This configuration
is depicted in the right part of Fig.~\ref{fig:USP}.

The open string spectrum is nicely encoded in the partition functions and
it can be worked out using standard methods. We choose the
type I string description with discrete value of $B_{ab}$,
in which the cylinder and the Moebius amplitudes are given by
\begin{equation}\label{oplus1}
  \begin{split}
  \mathcal{A} &= \frac{N^2}{2} \int_0^{\infty} \frac{d \tau_2}{\tau_2^5}
   \frac{V_8-S_8}{\eta^8} \big(\tfrac{i}{2}\tau_2\big)
   \big(P_{m_1}+ P_{m_1+\frac{1}{2}}\big)\big(P_{m_2}+
   P_{m_2+\frac{1}{2}}\big) \ , \\
\mathcal{M} &= \frac{N}{2} \int_0^{\infty} \frac{d \tau_2}{\tau_2^5} \frac{{\hat V}_8-{\hat S}_8}{{\hat \eta}^8} (\tfrac{i}{2}\tau_2+\tfrac{1}{2}) 
   \big[\big(P_{m_1}- P_{m_1+\frac{1}{2}}\big) P_{m_2}
   - \big(P_{m_1}+ P_{m_1+\frac{1}{2}}\big) P_{m_2+\frac{1}{2}}\big] \ .
\end{split}
 \end{equation}
Here $P_{m_i+a}$ are the momentum sums
 \begin{equation}\label{psum}
   P_{m_i+a} = \sum_{m_i} q^{\pi\alpha' (m_i+a)^2/R_i^2}\ , \quad
   q=e^{2\pi i\tau}\ ,
 \end{equation}
where $R_i$ are the radii of the type I torus.
Without the string excitations one obtains the open-string partition
function
\begin{equation}\label{partitionFT}
\begin{split}
  \mathcal{A} + \mathcal{M} = \int_0^{\infty} \frac{d \tau_2}{\tau_2^5}
   & \left(\frac{V_{8}-S_{8}}{\eta^8}\right)_0 \left(\frac{N(N+1)}{2}
   P_{m_1}P_{m_2}\right.\\
   &\left.+ \frac{N(N-1)}{2}\big(P_{m_1}P_{m_2+\frac{1}{2}}+
   P_{m_1+\frac{1}{2}}P_{m_2} + P_{m_1+\frac{1}{2}}P_{m_2+\frac{1}{2}}\big)\right) \ . 
\end{split}
 \end{equation}
The 8d spectrum is supersymmetric and corresponds to the gauge group $\text{USp}(N)$, with $N=16$ fixed by the RR tadpole conditions.
The field theory part of the spectrum can be read off from Eq.~\eqref{partitionFT}.
It contains massless states and several massive towers of the 2-torus:
\begin{itemize}
\item one massless gauge multiplet and KK towers
  in the {\it symmetric} (adjoint) representation of $\text{USp}(16)$,
  with masses 
\begin{equation}
M_1^2 = \frac{m_1^2}{R_1^2}+ \frac{m_2^2}{R_2^2}  \ ,\label{oplus2} 
\end{equation}
\item three massive gauge multiplets and KK towers in the {\it antisymmetric} representation of $\text{USp}(16)$, with masses 
\begin{equation}
M_2^2 = \frac{(m_1+\frac{1}{2})^2}{R_1^2}+ \frac{m_2^2}{R_2^2} \ , \  
M_3^2 = \frac{m_1^2}{R_1^2}+ \frac{(m_2+\frac{1}{2})^2}{R_2^2} \ , \ M_4^2 = \frac{(m_1+\frac{1}{2})^2}{R_1^2}+ \frac{(m_2+\frac{1}{2})^2}{R_2^2} \ , \label{oplus3} 
\end{equation}
\end{itemize}
where $R_1, R_2$ are the two radii of the type I torus.
It is remarkable that this spectrum agrees with
\eqref{massesFT}, the mass spectrum after tachyon
condensation obtained in Section~\ref{sec:massive} where, for simplicity, a
square torus was chosen ($R_1=R_2=R$).

The string vacuum just described is the only one preserving
supersymmetry in 8d with gauge group
$\text{USp}(16)$. It is therefore the only candidate for the endpoint
of tachyon condensation in the magnetized U(16) 8d type I
string. Hence, we are led to conjecture that the two string vacua are
equivalent. However, since the rank of the gauge group is divided by
two, the number of branes is halved in the process. In the T-dual
picture, we need to understand the surprising connection between
tachyon condensation and the conversion $\text{O}7_- \to \text{O}7_+$.  While we
do not have a full proof of the equivalence between the two string
vacua, the following facts appear to support our conjecture.

Consider  type I string with magnetized D9-branes wrapped on a torus
$T^2$ with coordinates $y_1,y_2$, and a magnetic field along a
Cartan generator commuting with SU(16). After T-duality on $y_2$, 
the magnetized D9-branes become a stack of D8-branes making an angle
$\tan \theta = k R'_2/R_1$ with the O8-planes, where $R'_2$ is the
T-dual radius. The tachyons in the antisymmetric representation of
$\text{U}(16)$ are stretched between the brane with wrapping numbers
$(1,k)$ and its image $(1,-k)$. 
The corresponding wave functions are exponentially localized at their
$I = 2 k$ intersections, one at the origin and the others depending on
$k$ (as an example, for $k=1$ there is a second intersection
in the middle of the torus, see the left part of Fig.~\ref{fig:USP}).
Suppose the tachyon is condensing at the origin. Tachyon condensation
corresponds to brane recombination. Here, the two cycles of the branes
and their images recombine  into a single factorizable
cycle $(2,0)$,  
which is parallel to the cycle wrapped by the O8-plane at the origin $y_2=0$. The corresponding D8-branes have gauge group $\text{USp}(16)$, as already discussed.  
The halving of the rank of the gauge group can be related to the resulting wrapping number  $(2,0)$, which doubles the length of the recombined D8 branes.
Since half of the D8-branes disappear in the process, presumably the nature of the O8-plane is also changed, in a way that cannot be captured in perturbation theory\footnote{Taking into account the resulting gauge group $\text{USp}(16)$, the torus in type IIA orientifold should become tilted, and the D8-branes and the O8-plane should acquire the wrapping numbers
$(2,-1)$ \cite{Blumenhagen:2000ea, Bachas:2008jv}.}.  
If one now T-dualizes also the $y_1$-coordinate, one ends  up with
D7-branes localized at the origin of the dual torus
$(y_1,y_2)=(0,0)$, with gauge group $\text{USp}(16)$.  The same
T-dualities for the $\text{SO}(32)$ type I string (with no magnetic
field) would of course lead to D7-branes with gauge group $\text{SO}(32)$. 

To shed some light on the conversion of one O7-plane, it is simpler to use the other, singular limit in which the magnetized D9-branes
are described by the D9-D7 brane system, as described in
  Section~\ref{sec:d9d7}. In our case, we start with
16 non-BPS D7-branes in $\text{SO}(32)$ type I string, with gauge group $\text{U}(16)$. As described in the previous section, there are two complex tachyons in the 
spectrum, in the representations $({\bf 32,16})$ and   $({\bf 1,120})$ of $\text{SO}(32) \times \text{U}(16)$. After two T-dualities in $y_1,y_2$, we obtain a non-BPS D9-brane with gauge group $\text{U}(16)$ and a BPS D7-brane
with gauge group $\text{SO}(32)$, located at the origin of the
torus. There are also the standard four $\text{O}7_-$-planes of the type IIB
orientifold with the T-dual orientifold projection $\Omega' = \Omega
\Pi (-1)^F $,  where $\Pi$ is the reflection $(y_1,y_2) \to
(-y_1,-y_2)$ and $(-1)^F$ is the spacetime fermion number. 
The condensation of the bi-fundamental tachyon  $({\bf 32,16})$ breaks the gauge group to the diagonal $\text{U}(16)$. The condensation of the tachyon in the antisymmetric representation $({\bf 1,120})$ breaks the 
gauge group to  USp(16) and halves the number of branes. What happens to the other eight branes, which disappear at the origin at the torus? Checking the charges/tension of these objects, one notices the suggestive equality
\begin{equation}  
Q_{\text{O}7_+} = Q_{\text{O}7_-} + 8 Q_{\text{D}7}   \ . \label{oplus3}
 \end{equation}
This leads to the conjecture that the eight branes disappearing after tachyon condensation form a bound state with the coincident $\text{O}7_-$-brane, the resulting object having precisely 
the charge and tension of an  $\text{O}7_+$-brane! The orientifold
conversion can be compared to the 
  $Q_{\text{O}3_{+}} = Q_{\text{O}3_{-}} + 1/2 \, Q_{\text{D}3}$ needed in S-duality.  While the $1/2$ stack (or rigid) D3 has no position
moduli and, intuitively, can generate a bound state with the original
$\text{O}3_{-}$, in our case the 8 $\text{D}7$-branes probably form a
  bound state with the $\text{O}7_{-}$-plane due to the tachyon condensation, and the 
corresponding open-string degrees of freedom become massive in the
process. It would clearly be of great interest to find further
evidence for this equivalence. However, this appears to be the only logical interpretation of the tachyon condensation in the context we discussed.  
Finally, notice that  the results of Section~\ref{sec:symmetry} exclude the interpretation of a $\text{USp}(16)$ type I vacuum in terms of  non-Abelian Wilson lines .


\section{Conclusions and open questions}
\label{sec:conclusion}

Understanding the vacuum structure of intersecting
  D-brane models and their T-dual magnetic compactifications is of
  crucial importance for potential phenomenological applications. In
  the interesting class of orientifold compactifications with chiral
  fermions, Wilson lines have generally been used to remove
  tachyons. Since quantum corrections have been shown to
  prevent local minima for Wilson lines, it is important to study
  tachyon condensation for vacua without Wilson lines.
  
The goal of this paper has been to analyze the simplest case of
tachyon condensation which occurs in eight-dimensional type I string
theory with magnetized branes wrapping a torus.  This setup is T-dual
to type IIA orientifolds with intersecting D8-branes. We have analyzed
in detail the simplest case where in the type IIA description all 16
D8-branes are rotated by the same angle, leading to the gauge group
$\text{U}(16)$ in the open string spectrum.
In this case a tachyon occurs at the intersection of the D8-branes
with their images. We have shown that after tachyon condensation the
resulting gauge group becomes $\text{USp}(16)$. By analyzing the field
theory condensation process, we have argued that the final string theory
should be equivalent to the unique known $\text{USp}(16)$ superstring
in eight dimensions. The latter has the peculiar feature
that, in a T-dual version with D7- and O7-planes, one orientifold
plane is of exotic type $\text{O}7_+$, whereas the three others are
standard $\text{O}7_-$-planes. We argue that tachyon condensation
induces dynamically an $\text{O}7_- \to \text{O}7_+$ conversion, by generating a
bound state of the $\text{O}7_-$-plane with half of the D7-branes that naively disappear in the condensation process. The result
is supported by various arguments and in particular by the equality 
\begin{equation}  
Q_{\text{O}7_+} =  Q_{\text{O}7_-} + 8 Q_{\text{D}7}   \nonumber\ , \label{concl1}
\end{equation}
yielding a consistent picture of a type IIB orientifold.

The stable vacuum after tachyon condensation is supersymmetric and has
no chiral fermions. It remains a challenge to find intersecting
D-brane models with broken supersymmetry and chiral fermions.


\section*{Acknowledgments}
We thank Costas Bachas, Markus Dierigl, Jihad Mourad, Augusto Sagnotti
and Timo Weigand for valuable
discussions and correspondence. E.D. was supported in part by the ``Agence Nationale de
la Recherche" (ANR). Y.T. is supported in part by Grants-in-Aid for
JSPS Overseas Research Fellow (No.~18J60383) from the Ministry of
Education, Culture, Sports, Science and Technology in Japan, 
and in part by Scuola Normale, by INFN (IS GSS-Pi) and by the MIUR-PRIN contract 2017CC72MK\_003.

\appendix

\section{Flux quantization}
\begin{figure}[t]
\begin{center} 
\includegraphics[width = 0.4\textwidth]{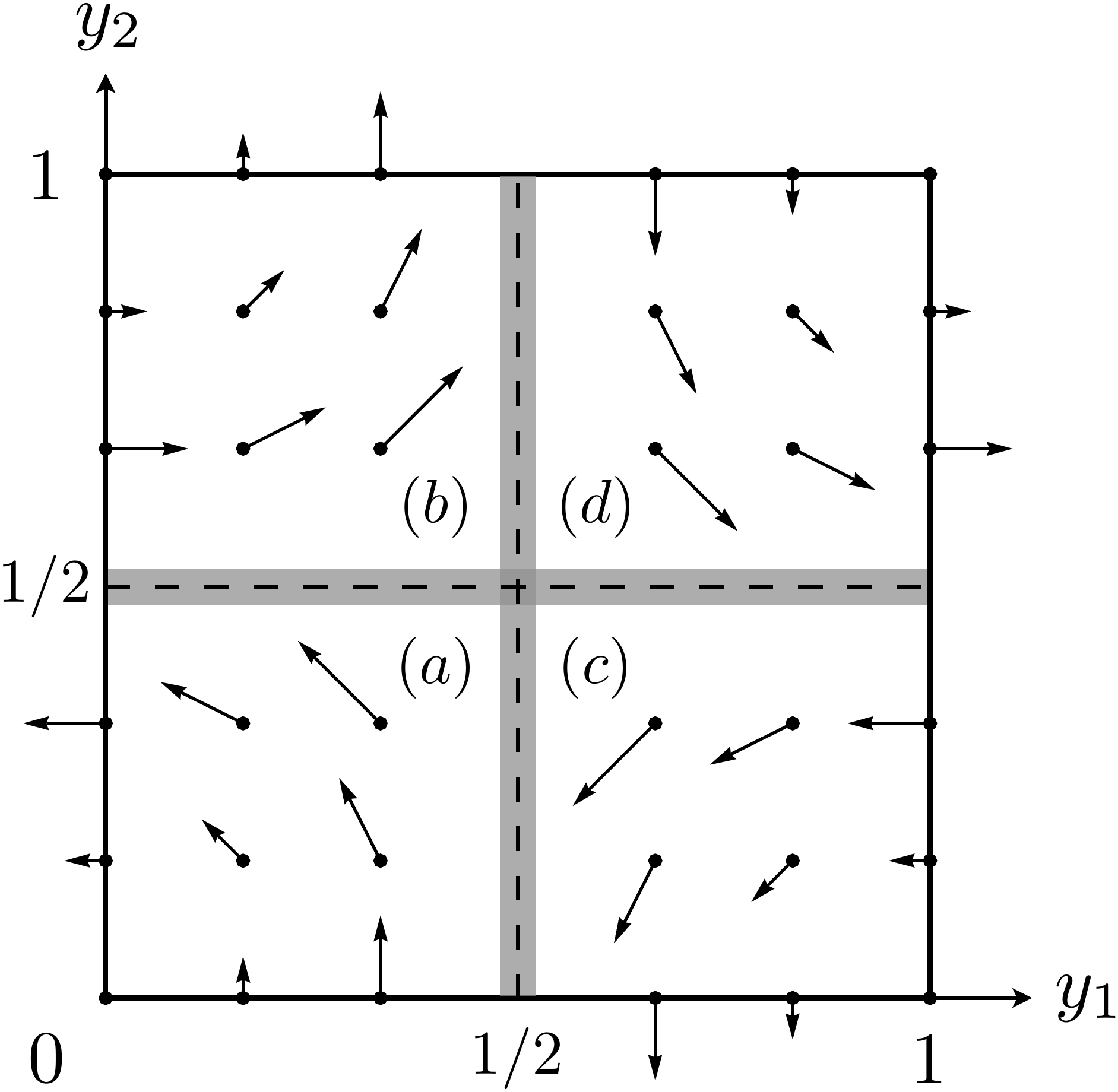}
\end{center}
\caption{Vector potential for constant magnetic flux on a torus with
  four patches.}
\label{fig:symmetricgauge}
\end{figure}
The torus $T^2$ is not simply connected so that vector fields and matter
fields are represented by fibre bundles. The case of a monopole field on a
sphere with constant flux density has been thoroughly discussed in
Ref.~\cite{Wu:1975es}, and homogeneous magnetic fields on a torus have
been considered in Refs.~\cite{Abouelsaood:1986gd,Buchmuller:2018lkz}. 
In these papers the vector potential for constant magnetic flux was
chosen in Landau gauge so that the torus could be covered with two
patches. Another convenient choice is the symmetric gauge \cite{Cremades:2004wa}
which requires four patches, for instance the ones shown in
Fig.~\ref{fig:symmetricgauge}. Here the vector potential is given by
\begin{align}\label{fluxfield}
\renewcommand{\arraystretch}{1.4}
  \begin{array}{lllll}
(a) \quad & A_1 = -\frac{f}{2} y_2\ , & A_2 = \frac{f}{2} y_1\ ,
  &\qquad 0 \leq y_1 < \frac{1}{2}\ , & 0 \leq y_2 < \frac{1}{2} \\
(b) \quad & A_1 = -\frac{f}{2} (y_2 - 1) \ , & A_2 = \frac{f}{2} y_1 \ ,
  & \qquad 0 \leq y_1 < \frac{1}{2}\ , & \frac{1}{2} \leq y_2 < 1 \\
(c) \quad & A_1 = -\frac{f}{2} y_2\ , & A_2 = \frac{f}{2} (y_1-1)\ ,
  &\qquad \frac{1}{2} \leq y_1 < 1\ , & 0 \leq y_2 < \frac{1}{2} \\
(d) \quad & A_1 = -\frac{f}{2} (y_2 - 1) \ , & A_2 = \frac{f}{2} (y_1-1) \ ,
  & \qquad \frac{1}{2} \leq y_1 < 1\ , & \frac{1}{2} \leq y_2 < 1 \ ,
\end{array}
\end{align}
leading to the constant magnetic field
$F = dA = f v$, with $v = dy_1\wedge dy_2$. Each patch overlaps with
two other patches at two boundaries each, see Fig.~\ref{fig:symmetricgauge}.

The transition function that
relates fields at the same point in the two patches is given by \cite{Wu:1975es}
\begin{equation}\label{overlap}
\begin{split}
\phi_f &= S_{fe} \phi_e\,, \quad S_{fe} = e^{i\Lambda_{fe}}\,, \quad
S_{fe} = S^{-1}_{ef}\,, \\
A_m^f &= A_m^e + \frac{i}{g} S^{-1}_{fe} \partial_m S_{fe}
= A_m^{e} - \frac{1}{g} \partial_m \Lambda_{fe}\,.
\end{split}
\end{equation}
Given the vector fields \eqref{fluxfield} in the patches $(a)$, $(b)$,
$(c)$ and $(d)$ this yields the transition functions
\begin{equation}
\Lambda_{ba}(y_1,1/2) = \Lambda_{dc}(y_1,1/2)= -\frac{gf}{2} y_1\ , \quad
\Lambda_{ca}(1/2,y_2) = \Lambda_{db}(1/2,y_2) = \frac{gf}{2} y_2\ , 
\end{equation}
The consistency condition at the overlap of all four patches,
\begin{equation}
\Lambda_{ac}(1/2,1/2)\Lambda_{cd}(1/2,1/2)
\Lambda_{db}(1/2,1/2)\Lambda_{ba}(1/2,1/2)
= e^{-igf} =1\ ,
\end{equation}
implies flux quantization, $gf = 2\pi M$, $M\in \mathbb{Z}$. The
transition functions at the boundaries of the fundamental domain of
the torus are trivial.

Starting at $y_2 = 1/2$ in patch $(b)$ and going around the torus in
$y_2$-direction via patch $(b)$ and patch $(a)$ until $y_2 = 1/2$ in patch $(a)$,
the vector field changes from $A_1 = f/4$ to $A_1 = -f/4$. 
This necessitates a non-trivial transition
function $S_{ba}$, which in the literature on magnetized tori 
usually occurs as twisted boundary condition (see, for example \cite{Cremades:2004wa}),
\begin{equation}
\phi(y_1,y_2 + 1) = S_{ba}^{-1}(y_1,1/2) \phi(y_1,y_2) = e^{iqfy_1/2}
  \phi(y_1,y_2)\ .
\end{equation}
Analogously, one obtains for a lattice translation in $y_1$-direction,
\begin{equation}
\phi(y_1+1,y_2) = S_{ca}^{-1}(1/2,y_2) \phi(y_1,y_2) = e^{-iqfy_2/2}
  \phi(y_1,y_2)\ .
\end{equation}
This means that the twist factors are given by the transition
functions on the torus in $y_1$- and $y_2$-direction, respectively.


\bibliography{tachyon}
\bibliographystyle{unsrt}

\end{document}